\documentclass[aps,prd,nofootinbib,superscriptaddress,preprint,eqsecnum,showkeys,showpacs,preprintnumbers]{revtex4}
\usepackage{color}
\usepackage{slashed}
\usepackage{graphics}
\usepackage{graphicx}
\usepackage{dcolumn}
\usepackage{subfigure}
\usepackage{mathrsfs}
\usepackage{bm}
\usepackage{amsmath,amssymb,epsfig}
\usepackage{float}
\allowdisplaybreaks[3]

\begin{document}
\newcommand{\bea}{\begin{eqnarray*}}
\newcommand{\eea}{\end{eqnarray*}}
\newcommand{\bean}{\begin{eqnarray}}
\newcommand{\eean}{\end{eqnarray}}
\newcommand{\eqs}[1]{Eqs. (\ref{#1})}
\newcommand{\eq}[1]{Eq. (\ref{#1})}
\newcommand{\meq}[1]{(\ref{#1})}
\newcommand{\grad}{\nabla}
\newcommand{\eqn}{&=&}
\newcommand{\non}{\nonumber \\}
\newcommand{\hsp}{\hspace{0.1mm}}
\newcommand{\pp}{\partial}

\newcommand{\vvr}{\mbox{\boldmath${r}$}}
\newcommand{\vp}{\mbox{\boldmath${P}$}}
\newcommand{\ve}{\mbox{\boldmath${e}$}}
\newcommand{\vchi}{\mbox{\boldmath${\chi}$}}
\newcommand{\vS}{\mbox{\boldmath${S}$}}
\newcommand{\vs}{\mbox{\boldmath${s}$}}

\title{Self-consistent effective-one-body theory for spinning   binaries  based on  post-Minkowskian approximation}

\author{Jiliang {Jing}\footnote{ jljing@hunnu.edu.cn}}
 \affiliation{Department of Physics, Key Laboratory of Low Dimensional Quantum Structures and Quantum Control of Ministry of Education, and Synergetic Innovation
Center for Quantum Effects and Applications, Hunan Normal
University, Changsha, Hunan 410081, P. R. China}
\affiliation{Center for Gravitation and Cosmology, College of Physical Science and Technology, Yangzhou University, Yangzhou 225009, P. R. China}

\author{Weike Deng}
\affiliation{Department
of Physics, Key Laboratory of Low Dimensional Quantum Structures and
Quantum Control of Ministry of Education, and Synergetic Innovation
Center for Quantum Effects and Applications, Hunan Normal
University, Changsha, Hunan 410081, P. R. China}

\author{Sheng Long}
 \affiliation{Department
of Physics, Key Laboratory of Low Dimensional Quantum Structures and
Quantum Control of Ministry of Education, and Synergetic Innovation
Center for Quantum Effects and Applications, Hunan Normal
University, Changsha, Hunan 410081, P. R. China}

\author{Jieci Wang\footnote{ jcwang@hunnu.edu.cn}} 
\affiliation{Department
of Physics, Key Laboratory of Low Dimensional Quantum Structures and
Quantum Control of Ministry of Education, and Synergetic Innovation
Center for Quantum Effects and Applications, Hunan Normal
University, Changsha, Hunan 410081, P. R. China}


\begin{abstract}

This paper extends the research on the self-consistent effective-one-body theory of a real spinless two-body system based on the post-Minkowskian approximation (Science China, 65, 100411, (2022)) to the case of a binary system for the spinning black holes. An effective rotating metric and an improved Hamiltonian for the spinning black hole binaries were constructed. The decoupled equation for the null tetrad component of the gravitational perturbed Weyl tensor $\psi^B_{4}$ in the effective rotating spacetime is found with the help of the gauge transform characteristics of the Weyl tensors. The decoupled equation is then separated between radial and angular variables in the slowly rotating background spacetime, and a formal solution of $\psi^B_{4}$ is obtained. On this basis, the formal expressions of the radiation reaction force and the waveform for the ``plus'' and ``cross'' modes of the gravitational wave are presented. These results, obtained in the same effective spacetime, constitute a self-consistent effective-one-body theory for the spinning black hole binaries based on the post-Minkowskian approximation.

\end{abstract}


\maketitle

\newpage
\tableofcontents
\newpage

 \section{Introduction}

The gravitational waves (GWs) \cite{einstein18,BonVanMet62,Sac62, PerAldDel98,Fang,Jing2021} generated by coalescing compact object binary systems are the most promising sources for the laser interferometer gravitational wave detectors, such as LIGO, Virgo, LISA, KAGRA, Taiji, and TianQin~\cite{Barish,Waldman,Acernese,schutz_lisa}. The gravitational waveform template plays a central role in the detection of the GW events generated by coalescing compact object binary systems.
Tens of thousands of waveform templates may be needed to extract the GW signal from the noise by means of the matched filtering techniques. The high computational cost and the large parameter space for spinning binary black holes show that it will be impractical for numerical relativity alone to provide a gravitational waveform template bank \cite{Pan3}.

To construct the gravitational waveform template for the gravitational radiation generated by coalescing compact object binary systems, the key point is to study the late dynamical evolution of the system. Buonanno and Damour \cite{Damour1999} introduced the effective-one-body (EOB) theory based on the post-Newtonian (PN) approximation. The EOB theory is a novel approach to studying the general gravitational radiation generated by coalescing compact object binary systems. It \cite{Damour1999,Damour2000(2),Damour20091, Damour2000,Damour2001} presents a theoretical framework that yields a resumed analytical description of the gravitational dynamics of two black holes and becomes the basis of the computation of many gravitational waveform templates \cite{Taracchini,Bohe}, which have been used in the data analysis of the gravitational wave signals detected by the LIGO and Virgo interferometers \cite{Abbott2016,Abbott2016(2),Abbott2017,Abbott2017(2)}.

In 2016, Damour~\cite{Damour2016} developed another EOB theory with a post-Minkowskian (PM) approximation to release the assumption that $v/c$ should be a small quantity in the EOB theory based on the PN approximation. Since then, the correlational research has attracted great attention \cite{Damour2017, Damour2018,Damour2018new,Antonelli2019,Damour2019, Damour2020,HeLin2016, Jing2019,Blanchet2018, Cheung2018,Vines2019,Cristofoli2019, Collado2019, Bern2019,Bern20192, Plefka2019,BiniDamour2020,Cheung2020}. Along this way, a self-consistent effective-one-body (SCEOB) theory \cite{Jing} was set up for the real spinless two-body system in PM approximation, which can be applied to the study of the dynamics and the gravitational wave emission of coalescing nonspinning black holes with two mass parameters $(m_1, m_2)$.
However, in general, a black hole has mass and spin. That is, a real spin black hole binary depends on the mass and spin parameters $(m_1,\boldsymbol{S}_1, m_2, \boldsymbol{S}_2)$. Therefore, the next important step is to extend the study of the SCEOB theory for the real spinless two-body system \cite{Jing} to the case of a binary system of spinning black holes.

For the real spin two-body system, the basic idea of the EOB theory is to map the conservative dynamics of two compact objects with masses $m_1$ and $m_2$ and spins $\vS_1$ and $\vS_2$ into the dynamics of an effective particle with mass $m_0 = m_1 m_2/(m_1 + m_2) $ and spin $\vS_*$ orbits around a massive black hole described by an effective rotating metric with mass parameter $M_0 = m_1 + m_2$ and rotational parameter $a$ \cite{Barausse}.
Thus, the first step in constructing the SCEOB theory for the binary system of spinning black holes is to find the effective rotating metric. Based on the effective metric \cite{Jing} of a real spinless two-body system for the EOB theory in the PM approximation, an effective rotating metric for a real spin two-body system will be built in this study by means of the approach of constructing an effective rotating metric presented by Damour or Barausse et al. \cite{Barausse,Damour}.

The second step is to find an improved EOB Hamiltonian.
The EOB Hamiltonian that includes spin effects has been studied extensively ~\cite{Damour2001,Buonanno2006,DamourH,BarausseH,BarausseH1,Damour2009,Pan2}.
Damour, Jaranowski, and Schafer \cite{DamourH} presented an EOB Hamiltonian to explore the calibration of numerical relativity waveforms of spinning black holes within the EOB formalism. Later,
in the effective Kerr background spacetime, Barausse, Racine, and Buonanno \cite{BarausseH1} constructed an improved EOB Hamiltonian for a spinning test particle at a linear order of the particle spin. We use the Barausse--Racine--Buonanno proposal to construct an expression of an improved SCEOB Hamiltonian for a spinning test particle that orbits around a massive rotating black hole described by the effective rotating metric.

As a coalescing compact object binary system generates the GWs, the energy of the binary system will be lost, giving rise to the radiation reaction force (RRF). The RRF for these modes of the GWs must be determined to detect the ``plus'' and ``cross'' modes of the GWs.
Therefore, the third step is to determine the RRF by means of the gravitational wave energy flux for the ``plus'' and ``cross'' modes of the GWs. In order to do that, the decoupled equation for the null tetrad component of the perturbed Weyl tensor $\psi^B_{4}$ must be determined since it relates to the ``plus'' and ``cross'' modes of the GWs as $\psi^B_4=\frac{1}{2}(\ddot h_{+}-i\ddot h_{\times})$ at infinity.
Before our studies, for the gravitational perturbation, the decoupled equation of $\psi^B_{4}$ was only obtained in the Schwarzschild and Kerr spacetimes. In Ref. \cite{Jing1}, by dividing the perturbation part of the metric into the odd and even parities, we found the decoupled equations
of the tetrad component of the perturbed Weyl tensor $\psi^B_{4}$ for the spinless binaries by taking the Regge--Wheeler gauge \cite{Thompson} in the effective metric. In 
Ref. \cite{Jing}, by means of the gauge transform property of the tetrad components of the perturbed Weyl tensors, we then obtained a new decoupled equation with separable variables of $\psi^{B}_{4}$ by taking a gauge in which $\psi_{1}^{B}$  and $\psi_{3}^{B}$ vanish in the effective metric.
Compared with the two theories, we know that the computational effort for the RRF and waveform in the latter theory will be tremendously reduced.
Especially, we know that the study for the spinless binaries in Ref. \cite{Jing} to the spinning black holes can be extended by checking all conditions to obtain the decoupled equation of $\psi^B_{4}$.
That is, we can find the decoupled equation of the tetrad component of the perturbed Weyl tensor $\psi^B_{4}$ in the effective rotating metric and separate the decoupled equation in the radial and angular parts in a slowly rotating background and then obtain the RRF expression by means of the flux for the ``plus" and ``cross" modes of the GW energy emitted to infinity in the effective rotating spacetime.

The last step is to construct the waveform for the coalescing compact object binary systems. By comparing the tetrad component of the perturbed Weyl tensor $\psi^B_4$ with the formula for the waveform \cite{Kidder} of the ``plus" and ``cross" modes of the GWs, we can read out the waveform $h^{l m}$ easily, which is also based on the effective rotating spacetime. Furthermore, we can use Damour--Nagar--Pan's proposals to improve the waveform \cite{Damour2007,Damour2009,Pan2,Pan3}. That is, based on the obtained waveform, the improved multipolar waveforms can be built as a product of the leading Minkowskian order term, the relativistic conserved energy or angular momentum of the effective moving source, an infinite number of leading logarithms entering the tail effect, an additional phase correction, the remaining PM effects, and a non-quasicircular effect.

 The rest of the paper is organized as follows. Section II derives an effective rotating metric and an improved Hamiltonian of the EOB theory for a real spin two-body system. Section III presents a decoupled equation with a separation of variables and a formal solution of $\psi^B_4$ in the effective rotating spacetime. The expressions of the RRF and the waveform for ``plus" and ``cross" modes of the GWs by means of $\psi^B_4$ are then presented, and an SCEOB theory for the spin binaries based on the PM approximation is set up. The last section presents the final conclusions and discussions.

\section{Effective rotating metric and Hamiltonian of the EOB theory for a real spin two-body system}

It is well known that the theoretical model of gravitational radiation is the basis of the gravitational waveform template, and the key point to constructing the theoretical model is to study the late dynamical evolution of a coalescing binary system of compact objects.
The EOB theory is a novel approach to studying the two-body dynamics of compact objects with the goal of extending the analytical templates throughout the last stages of inspiral, plunge, merger, and ringdown.
The dynamical evolution of the EOB theory for a real spin two-body system can be described by the Hamilton equations \cite{Damour2001,Taracchini1}:
\begin{subequations}\label{HEq}
  \begin{eqnarray}
    \frac{d\vvr}{d\hat{t}}&=&\{\vvr,\hat{H}_{\text{real}}\}=\frac{\partial \hat{H}_{\text{real} }}{\partial \vp}\,,\\\label{EOM2}
    \frac{d\vp}{d\hat{t}}&=&\{\vp,\hat{H}_{\text{real}}\}+\hat{\bm{\mathcal{F}}}=-\frac{\partial \hat{H}_{\text{real}}}{\partial \vvr}
    +\hat{\bm{\mathcal{F}}}\,,\\\label{EOM3}
    \frac{d\vS_1}{d\hat{t}} &=& \{\vS_1, \nu \hat{H}_{\text{real}} \} = \nu \frac{\partial \hat{H}_{\text{real}}}{\partial \vS_1} \times \vS_1 \,,\\\label{EOM4}
    \frac{d\vS_2}{d\hat{t}} &=& \{\vS_2, \nu \hat{H}_{\text{real}} \} = \nu \frac{\partial \hat{H}_{\text{real}}}{\partial \vS_2} \times \vS_2 \,,
  \end{eqnarray}
\end{subequations}
where $\hat{t}\equiv t/M_0$,
$\hat{H}_{\text{real}}=H_{\text{real}}/m_0$ is the reduced EOB Hamiltonian \cite{BarausseH,BarausseH1,Barausse}, and
$\hat{\bm{\mathcal{F}}}=\bm{\mathcal{F}}/m_0$ is the reduced RRF \cite{Buonanno2006}, with $M_0=m_1+m_2$, $m_0=m_1m_2/(m_1+m_2)$, and $\nu=\frac{m_0}{M_0}=\frac{m_1 \, m_2}{(m_1+m_2)^2}$.

%
To get the explicit expression of $\hat{H}_{\text{real}}$ in the Hamilton equations (\ref{HEq}), we should know the effective metric $g_{\mu\nu}^{\text{eff}}$ and the energy map between the EOB and real two-body systems. In this section, we first built up an effective rotating metric of the EOB theory for the real spin two-body system based on the effective metric of the EOB theory for the real spinless two-body system and then found an improved EOB Hamiltonian.

\subsection{Effective rotating metric of the EOB theory for the real spin two-body system}

The effective metric of a real spinless two-body system for the EOB theory, in the PM approximation, can be expressed as \cite{Jing}
\begin{eqnarray}
ds_{\text{eff}}^2=\frac{\Delta^0_{r}}{r^2} dt^2-\frac{r^2}{\Delta^0_{r}}dr^2- r^2(d\theta^2+\sin^2\theta d\varphi^2),\label{Mmetric}
\end{eqnarray}
with
\begin{eqnarray}
&& \Delta^0_{r}=r^2- 2 GM_0 r+\sum_{i=2}^\infty a_i \frac{(GM_0\big)^i}{r^{i-2}},
\end{eqnarray}
where $M_0$ is the mass parameter of the black hole and
\begin{eqnarray} \label{parameters}
a_2&=&\frac{3(1- \, \Gamma)(1-5 \, \gamma^2)}{\Gamma\, (3\, \gamma^2-1 )} ,\nonumber \\
a_3&=&\frac{3}{2 }\Big[\frac{3- 2\, \Gamma-3 (15 -8\, \Gamma ) \gamma^2+6 ( 25-16 \, \Gamma ) \gamma^4}{\Gamma\, (4\, \gamma^2-1 ) (3\, \gamma^2-1 )}-\frac{2 P_{30}}{(4 \gamma^2-1 )}\Big], \nonumber \\    && \cdot \cdot \cdot  \cdot  \cdot  \  \cdot .
\nonumber  \end{eqnarray}
in which $\gamma = \frac{1}{2} \frac{\mathcal{E}^2-m_1^2-m_2^2}{m_1m_2} $, $\Gamma = \frac{\mathcal{E}}{m_1+m_2}$, $\mathcal{E}$ is the energy of the real two-body system, and $
P_{30} = \frac{18\gamma^2-1}{2\,\Gamma^2}+\frac{8\,\nu\, (3+12\gamma^2-4\gamma^4)}{\Gamma^2\, \sqrt{|\gamma^2-1|}} \mbox{arcsinh}\sqrt{\frac{|\gamma-1|}{2}}
+\frac{\nu}{\Gamma^2}\big(1-\frac{103}{3}\gamma-18 \gamma^2-\frac{2}{3} \gamma^3+\frac{3 \,\Gamma\,(1-2\gamma^2)(1-5 \gamma^2)}{(1+\Gamma)(1+\gamma)}\big).
$

Based on the effective metric (\ref{Mmetric}), using the approach of constructing an effective rotating metric presented by Damour or Barausse et al. \cite{Barausse,Damour} \footnote{please note that, starting from the spinless metric (\ref{Mmetric}), both the approaches presented by Damour or Barausse et al. will give the same effective rotating metric (\ref{effmetric}).}, we find that the effective rotating metric for a real spin two-body system is described by
\begin{align}\label{effmetric}
 ds^{2}=g^{\text{eff}}_{\mu\nu}d x^\mu d x^\nu=\frac{\Delta_{r}-a^2\sin ^2\theta}{\Sigma}dt^{2}-\frac{\Sigma}{ \Delta_{r}} dr^{2}-\Sigma d\theta^{2}-\frac{\Lambda_{t} \sin^2\theta }{\Sigma} d\phi^{2}+\frac{2\omega_{j} \sin^2\theta}{ \Sigma}dt d\varphi,
\end{align}
with
\begin{align}
\nonumber & \Sigma=\overline{\rho} \overline{\rho}^{*}, \quad\overline{\rho}=r+i a \cos\theta, \quad \overline{\rho}^{*}=r-i a \cos\theta, \quad \Delta_{r}=\Delta^0_{r}+a^2, \\
\nonumber &\Lambda_{t}=\varpi^{4}-a^{2} \Delta_{r} \sin^{2}\theta, \quad\varpi=(r^{2}+a^{2})^{\frac{1}{2}},\quad\omega_{j}=a(a^{2}+r^{2}-\Delta_{r}),
\end{align}
where $a$ is the rotational parameter. The metric can be applied to any post-Minkowskian orders. It is easy to show that the effective spacetime (\ref{effmetric}) is still type $D$.

\subsection{ Hamiltonian of the EOB theory for the real spin two-body system}

The relationship between the energy of the EOB system $\mathcal{E}_0$ and that of a real spinless two-body system $\mathcal{E}$ is given by \cite{Jing}
\begin{eqnarray}\label{EMap}
\mathcal{E}_0&=\frac{\mathcal{E}^2-m_1^2-m_2^2}{2(m_1+m_2)}\label{energymap1}. \end{eqnarray}
After Damour \cite{Damour2001}, we assume that Eq. (\ref{energymap1}) still holds for the real spin two-body system.
By means of the effective rotating metric (\ref{effmetric}) and the energy map (\ref{EMap}), and taking the approach of constructing the EOB Hamiltonian presented by Damour and Barausse et al.,~\cite{Damour,DamourH,BarausseH,BarausseH1}, we can construct an improved EOB Hamiltonian for a spinning black hole described by the effective rotating metric (\ref{effmetric}).

As shown in Ref~\cite{BarausseH1}, in curved spacetime, the total effective Hamiltonian of a spinning test particle at a linear order in the particle's spin can be written as
\begin{equation} \label{HHH}
\bar{H}_{\text{eff}} [g_{\mu\nu}^{\text{eff}}]= \bar{H}_{\rm NS} + \bar{H}_{\rm S},
 \end{equation}
 with
\begin{eqnarray}
&&{\bar{H}}_{\rm NS} = \beta^i \, P_i + \alpha \sqrt{m_0^2 - \gamma^{ij}\,P_i\,P_j + {\cal Q}_4(P)}\,,
\label{HNS}\\
\label{HS}
&&{\bar{H}}_{\rm S} = \left(\beta^i\,F_i^K + F_t^K - \frac{\alpha \gamma^{ij}\,P_i\,F_j^K}{\sqrt{m_0^2 +
\gamma^{ij}P_iP_j}}\right)\,S_K \,,
\end{eqnarray}
where $\bar{H}_{\rm NS}$ is the Hamiltonian for a nonspinning particle in the effective rotating metric \cite{Damour2000(2)}, $\bar{H}_{\rm S}$ is the first-order Hamiltonian in the particle's spin approximation\footnote{We should note that the signature of the spacetime we used in this paper is different from that in Ref. \cite{BarausseH1}.}, ${\cal Q}_4(P)$ is a quartic term in the space momenta $P_i$ which was introduced in Ref.~\cite{Damour2000(2)},
$\alpha= \frac{1}{\sqrt{g_{\text{eff}}^{tt}}}\,,$ $ \beta^i = \frac{g_{\text{eff}}^{ti}}{g_{\text{eff}}^{tt}}\,,$ $
\gamma^{ij} =\Big( g_{\text{eff}}^{ij}-\frac{g_{\text{eff}}^{ti}\,g_{\text{eff}}^{tj}}{g_{\text{eff}}^{tt}}\Big)\,,
$ $S_K$ is the three-dimensional spin vector \cite{BarausseH}, and
$
F_\mu^K = \left(2E_{\mu TI}\,\frac{\bar{\omega}_J}{\bar{\omega}_T} + E_{\mu IJ}\right)\epsilon^{IJK},$
with
$
E_{\lambda\mu\nu} =  \frac{1}{2}\,\eta_{AB}\,\tilde{e}_\mu^A\,\tilde{e}_{\nu;\lambda}^{B}, $ $
\bar{\omega}_\mu=\bar{P}_\mu-m_0 \,\tilde{e}^{T}_\mu,$ $
 \bar{P}_i = P_i,\ $ $
\bar{P}_t = -\beta^i\,P_i-\alpha\, \sqrt{m_0^2 -\gamma^{ij}\,P_i\, P_j},$ $\label{bomegaT}
\bar{\omega}_T = \bar{\omega}_\mu\,\tilde{e}^\mu_{T}=\bar{P}_\mu
 \tilde{e}^\mu_{T}-m_0\,,\ $ and $
\bar{\omega}_I =\bar{\omega}_\mu\,
 \tilde{e}^\mu_{I}= \bar{P}_\mu\,\tilde{e}^\mu_{I}.
$

For the effective rotating metric (\ref{effmetric}), we take the reference tetrad as
  \begin{eqnarray}
 \tilde{e}^T_\mu &=& \delta^t_\mu \sqrt{\frac{\Delta_r \Sigma}{\Lambda_t }}\,,\ \ \
 \tilde{e}^1_\mu = \delta^r_\mu \sqrt{\frac{\Sigma}{\Delta_r }}\,, \nonumber \\
 \tilde{e}^2_\mu &=& \delta^\theta_\mu \sqrt{\Sigma}\,, \ \ \ \
 \tilde{e}^3_\mu = - \frac{\omega_{j} \sin\theta }{\sqrt{\Lambda_t \Sigma} } \delta^t_\mu +\delta^\phi_\mu \sin\theta \sqrt{\frac{\Lambda_t} {\Sigma}}\,.
 \end{eqnarray}
After tedious calculations, we find that the Hamiltonian (\ref{HHH}) for the effective spacetime (\ref{effmetric}) can be rewritten as
\begin{equation}\label{HH}
\bar{H}_{\text{eff}} [g_{\mu\nu}^{\text{eff}}] = \bar{H}_{\rm NS} + \mathcal{K}^I S_I\,,
\end{equation}
with
\begin{widetext}
\begin{eqnarray}
\mathcal{K}^1 &=& -\frac{\sqrt{\Delta_r }\cos\theta}{\Lambda_t^2\sqrt{\Sigma Q}(1+\sqrt{Q})\sin^2\theta}\Big[(1+\sqrt{Q})(\Delta_r \Sigma^2+ \varpi^4(\varpi^2-\Delta_r)) + a\,\omega_{j}\,\varpi^2 \sqrt{Q}\sin^2\theta\Big]\hat{P}_\phi \nonumber
\\
& +&\frac{\Delta_r (r\,\omega_{j}\,\Sigma + a\,\varpi^2\xi)\sin\theta}{\Lambda_t^{3/2}\Sigma^2\sqrt{Q}(1+\sqrt{Q})}\hat{P}_r\hat{P}_\theta + \frac{a^2 \omega_{j}\, \Delta_r \cos\theta \sin^2\theta}{\Lambda_t^{3/2} \Sigma \sqrt{Q}(1+\sqrt{Q})}\Big(1+\sqrt{Q} + \frac{2\Sigma\hat{P}_\phi^2}{\Lambda_t \sin^2\theta} + \frac{\Delta_r \hat{P}_r^2}{\Sigma} \Big), \nonumber \\ \nonumber \\
\mathcal{K}^2 &=& \bigg[\frac{\Delta_r (r\Sigma^2 - a^2\xi\,\sin^2\theta)}{\Lambda_t^2\sqrt{\Sigma Q}\sin\theta} - \frac{(\varpi^4\,\xi - r\,\omega_{j}^2\sin^2\theta)}{\Lambda_t^2\sqrt{\Sigma}(1+\sqrt{Q})\sin\theta}\bigg]\hat{P}_\phi + \frac{a^2 \omega_{j}\,\Delta_r ^{3/2}\cos\theta\sin^2\theta}{\Lambda_t^{3/2}\Sigma^2 \sqrt{Q}(1+\sqrt{Q})}\hat{P}_r\hat{P}_\theta \nonumber \\
& +& \frac{\sqrt{\Delta_r }(r\,\omega_{j}\,\Sigma +a\, \varpi^2\xi)\sin\theta}{\Lambda_t^{3/2}\Sigma \sqrt{Q}(1+ \sqrt{Q})}\Big(1 + \sqrt{Q} + \frac{2\Sigma}{\Lambda_t \sin^2\theta}\hat{P}_\phi^2 + \frac{1}{\Sigma}\hat{P}_\theta^2\Big)\,, \nonumber \\ \nonumber \\
\mathcal{K}^3 &=&  - \frac{a^2 \Delta_r \cos\theta\sin\theta}{(\Lambda_t\Sigma)^{3/2}\sqrt{Q}(1+\sqrt{Q})}\Big(\Lambda_t + \sqrt{Q}\Delta_r \Sigma \Big)\hat{P}_r \nonumber \\ &- &\frac{r\Lambda_t\Delta_r \Sigma +\sqrt{Q}\Big(\Lambda_t (r \Delta_r \Sigma- \varpi^2 \xi) +\omega_{j}\big(a \varpi^2\xi+r \omega_{j}\Sigma\big)\sin^2\theta\Big)}{\Lambda_t^{3/2}\Sigma^{5/2}\sqrt{Q}(1+\sqrt{Q})}\hat{P}_\theta \nonumber \\
& -&\frac{\sqrt{\Delta_r }}{\Lambda_t^2 \Sigma \sqrt{Q}(1+\sqrt{Q})}\bigg[a^2\omega_{j}\,\Delta_r \cos\theta\sin\theta \hat{P}_r + (r\,\omega_{j}\,\Sigma +a \varpi^2\xi)\hat{P}_\theta\bigg]\hat{P}_\phi \,,
\end{eqnarray}
\end{widetext}
where $\xi=a^2\,r\,\sin^2\theta-r\, \Delta_r+\frac{1}{2}\Sigma\,\Delta_r'$,
 $ Q = 1 +\frac{\Delta_r }{\Sigma} \hat{P}_r^2+\frac{1}{\Sigma} \hat{P}_\theta^2+\frac{\Sigma}{\Lambda_t \sin^2\theta} \hat{P}_\phi^2\,,
$
and $\hat{P}_i \equiv P_i/m_0$. Here and hereafter, a prime denotes derivation with respect to $r$.

Thus, by means of the energy map \eqref{EMap} and the total effective Hamiltonian (\ref{HH}), we find that the Hamiltonian can be expressed as $H_{\text{real}}=M_0 \sqrt{1+2\nu \left(\bar{H}_{\text{eff}} [g_{\mu\nu}^{\text{eff}}]/m_0 -1 \right)}\,$. Using the definitions of the reduced Hamiltonians $\hat{H}_{\text{real}}=H_{\text{real}}/m_0$ and $\hat{H}_{\text{eff}}[g_{\mu\nu}^{\text{eff}}]=\bar{H}_{\text{eff}} [g_{\mu\nu}^{\text{eff}}]/m_0$, we then find that an improved reduced EOB Hamiltonian which appeared in Eq (\ref{HEq}) is described by
\begin{eqnarray} \label{Heob}
\hat{H}_{\text{real}}=\frac{1}{\nu} \sqrt{1+2\nu \left(\hat{H}_{\text{eff}} [g_{\mu\nu}^{\text{eff}}]-1 \right)}\,,
\end{eqnarray}
which shows that the improved reduced Hamiltonian $\hat{H}_{\text{real}}$ is constructed in terms of the effective rotating metric.

\section{Radiation reaction force and waveform for ``plus" and ``cross" modes of gravitational wave }

The key step to determine the RRF $\bm{\mathcal{F}}$ and the waveform $h^{l m}$ for the ``plus" and ``cross" modes of the gravitational wave is to obtain the decoupled and variables separated equation for the null tetrad component of the perturbed Weyl tensor $\psi^B_{4}$ under the gravitational perturbation in the effective rotating spacetime. In this section, we first find the
decoupled equation of $\psi^B_{4}$ in the effective rotating spacetime. By separating the variables for the slowly rotating case, we then obtain the radial equation and its formal solution for $\psi^B_{4}$. At last, we present the formal expressions of the RRF and the waveform for the ``plus" and ``cross" modes of the gravitational wave in the effective rotating spacetime.

\subsection{Decoupled equation of $\psi^B_{4}$ for the gravitational perturbation in the effective rotating spacetime}

Taking the null tetrad for the effective rotating metric (\ref{effmetric}) as
\begin{align}
\nonumber &l^{\mu}=\Big\{\frac{r^{2}+a^{2}}{\Delta_{r}}, 1 ,0,\frac{a}{\Delta_{r}}\Big\}, \\
\nonumber &n^{\mu}=\Big\{\frac{r^{2}+a^{2}}{2 \overline{\rho} \overline{\rho}^{*}},-\frac{\Delta_{r}}{2\overline{\rho} \overline{\rho}^{*}},0,\frac{a}{2\overline{\rho} \overline{\rho}^{*}}\Big\}, \\
\nonumber &m^{\mu}=\frac{1}{\sqrt{2}\overline{\rho}}\Big\{ia\sin\theta,0,1,\frac{{i}}{\sin\theta}\Big\}, \\
 &\overline{m}^{\mu}=\frac{1}{\sqrt{2}\overline{\rho}^*}\Big\{-ia\sin\theta,0,1,-\frac{{i}}{\sin \theta}\Big\},
\end{align}
we find that all nonvanishing spin coefficients, the component of the trace-free Ricci tensor, and the component of the Weyl tensor in the effective spacetime (\ref{effmetric}) are given by
\begin{align}\label{spin}
&\nonumber \rho=-\frac{1}{\overline{\rho}^{*}}, \quad\mu=-\frac{\Delta_{r}}{2\overline{\rho}\overline{\rho}^{*2} }, \quad\gamma=\frac{\Delta_{r}^{\prime}}{4\overline{\rho}\overline{\rho}^{*}}+\mu, \quad\pi=\frac{i a\sin\theta}{\sqrt{2}\overline{\rho}^{*2} }, \\ &\nonumber \tau=-\frac{i a \sin\theta}{\sqrt{2}\overline{\rho}\overline{\rho}^{*}}, \quad\alpha=\pi-\frac{\cot\theta}{2\sqrt{2}\overline{\rho}^{*}} , \quad\beta=\frac{\cot\theta}{2\sqrt{2}\overline{\rho}}, \\
\nonumber &\phi_{11}=\frac{1}{8 \overline{\rho}^2 \overline{\rho}^{*2} }\Big[4\Delta_{r}-4 r \Delta_{r}^{\prime} + \overline{\rho}\overline{\rho}^{*} \Delta_{r}^{\prime\prime}+2 r^2 -5 a^2 - a^2 \cos^2(2\theta)\Big],\\
 &\psi_{2}=\frac{1}{12 \overline{\rho} \overline{\rho}^{*3} }\Big[12\Delta_{r}-6 \overline{\rho}^{*}\Delta_{r}^{\prime} + \overline{\rho}^{*2} \Delta_{r}^{\prime\prime}-2\Big(\overline{\rho}^{*2} +6 a (a+ i r \cos\theta)\Big)\Big],
 \end{align}

The equations in the Newman--Penrose formalism, which will be used to obtain the decoupled equation of $\psi_{4}$, are \cite{Teukolsky,Carmeli}
\begin{align}
 &\Delta\lambda-\overline{\delta}\nu=-(\mu+\overline{\mu})\lambda-(3\gamma-\overline{\gamma})\lambda+(3\alpha+\overline{\beta}+\pi-\overline{\tau})\nu-\psi_{4}, \label{Eq1}\\
 \nonumber&\overline{\delta} \psi_{3}-D\psi_{4}+\overline{\delta} \phi_{21}-\Delta \phi_{20}
 =3 \lambda \psi_{2}-2(\alpha+2 \pi) \psi_{3}+(4\epsilon-\rho) \psi_{4}-2 \nu \phi_{10}+2 \lambda \phi_{11}\nonumber \\ & \ \ \ \ +(2 \gamma-2 \overline{\gamma}+\overline{\mu}) \phi_{20}+2(\overline{\tau}-\alpha)\phi_{21}-\overline\sigma \phi_{22}, \label{Eq2} \\ &\Delta \psi_{3}-\delta \psi_{4}+\overline{\delta} \phi_{22}-\Delta \phi_{21} =3 \nu \psi_{2}-2(\gamma+2 \mu) \psi_{3}+(4 \beta-\tau) \psi_{4}-2 \nu \phi_{11}-\overline{\nu} \phi_{20}\nonumber \\ &\ \ \ \ +2 \lambda \phi_{12}+2(\gamma+\overline{\mu}) \phi_{21}+(\overline{\tau}-2 \overline{\beta}-2 \alpha) \phi_{22}.\label{Eq3}\end{align}

The gravitational perturbation in the background spacetime can be described by
\begin{align}\label{Pmetric}
  g_{\mu \nu}=g_{\mu \nu}^{A}+\varepsilon h_{\mu \nu}^{B}\;,
\end{align}
where the superscripts $A$ and $B$ denote the background and perturbation quantities, respectively, the background spacetime is taken as the effective metric $g_{\mu \nu}^{A}= g_{\mu \nu}^{\text{eff}}$, and $\varepsilon$ is a small quantity. Describing the gravitational perturbation (\ref{Pmetric}) by the Newman--Penrose formalism and keeping $\varepsilon$ up to the first order as we did in Ref. \cite{Jing}, from Eqs. (\ref{Eq1}), (\ref{Eq2}), and (\ref{Eq3}) we obtain
\begin{align}\label{Eq11}
 &\psi_{4}^{B}+(\Delta+3\gamma-\overline{\gamma}+\mu+\overline{\mu})\lambda^{B}-(\overline{\delta}+3\alpha+\overline{\beta}+\pi-\overline{\tau})\nu^{B}=0,\\
 &\nu^{B} (3\psi_{2}-2\phi_{11})-(\Delta+2\gamma+4\mu)\psi_{3}^{B}+(\delta+4\beta-\tau)\psi_{4}^{B}\nonumber \\&\ \ \ \ +(\Delta+2\gamma+2\overline{\mu})\phi_{21}^{B}-(\overline{\delta}+2\alpha+2 \overline{\beta}-\overline{\tau})\phi_{22}^{B}=0,\label{Eq22}\\
 &\lambda^{B}(3\psi_{2}+2\phi_{11})-(\bar\delta+2\alpha+4\pi)\psi_{3}^{B}+(D-\rho)\psi_{4}^{B}\nonumber \\&\ \ \ \ -(\overline{\delta}+2\alpha-2\overline{\tau})\phi_{21}^{B}+(\Delta+2\gamma -2\overline{\gamma}+\overline{\mu})\phi_{20}^{B}=0.\label{Eq33}\end{align}
Here and hereafter, for simplicity and clarity, we omit the superscript $A$.

 Chandrasekhar \cite{Chandrasekhar} pointed out that, for the linear perturbation described by Eq. (\ref{Pmetric}), $\psi_{0}^{B}$ and $\psi_{4}^{B}$ are gauge invariant, while the $\psi_{1}^{B}$ and $\psi_{3}^{B}$ are not. Therefore, taking a gauge in which $\psi_{1}^{B}$ and $\psi_{3}^{B}$ vanish, we find that, in the effective rotating spacetime, the decoupled equation of $\psi_{4}^{B}$ is described by
\begin{align}\label{decouple}
&\Big[\Big((\Delta+3 \gamma-\overline{\gamma}+\mu+\overline{\mu})-F_{1}\Big)(D-\rho)
-F_{2}\Big((\overline{\delta}+3 \alpha+\overline{\beta}+\pi-\overline{\tau})\nonumber \\& -F_{3} \Big)(\delta+4 \beta-\tau)-F_4\Big]\psi_{4}^{B}=T_4^B,
\end{align}
with
\begin{align}\label{T4}
\nonumber T_4^B&=\Big((\Delta+3 \gamma+\mu-\overline{\gamma}+\overline{\mu})-F_{1}\Big)\Big[(\overline{\delta}+2\alpha-2\overline{\tau})\phi_{21}^{B}-(\Delta+2 \gamma-2\overline{\gamma}+\overline{\mu})\phi_{20}^{B}\Big]\\
 &-F_{2}\Big((\overline{\delta}+3 \alpha+\overline{\beta}+\pi-\overline{\tau})-F_{3} \Big)\Big[(\overline{\delta}+2 \alpha+2 \overline{\beta}-\overline{\tau})\phi_{22}^{B}-(\Delta+2 \gamma+2 \overline{\mu}) \phi_{21}^{B}\Big],
\end{align}
where
\begin{align}
\nonumber &F_{1}=-\frac{\Delta_{r}}{2\overline{\rho} \overline{\rho}^{*}} \frac{\partial log[3\psi_{2}+2\phi_{11}]}{\partial r},\\
\nonumber &F_{2}=\frac{3\psi_{2}+2\phi_{11}}{3\psi_{2}-2\phi_{11}},\\\nonumber &F_{3}=\frac{1}{\sqrt{2}\overline{\rho}^* }\frac{\partial log[3\psi_{2}-2\phi_{11}]}{\partial \theta},\\
 &F_{4}=3\psi_{2}+2\phi_{11}.
\end{align}
To simplify the above equations, we introduce
\begin{align}
\nonumber &\mathscr{D}_{n}=\partial_{r}+\frac{i K}{\Delta_{r}}+n\frac{\Delta'_{r}}{\Delta_{r}}, \qquad\mathscr{D}_{n}^{\dagger}=\partial_{r}-\frac{iK}{\Delta_{r}}+n\frac{\Delta'_{r}}{\Delta_{r}}, \\
 &\mathscr{L}_{n}=\partial_{\theta}+Q+n\cot\theta, \qquad\mathscr{L}_{n}^{\dagger}=\partial_{\theta}-Q+n\cot\theta,
\end{align}
with
\begin{align}
 K=(r^2+a^2)\sigma^{\dagger} + m a , \qquad Q=\sigma^{\dagger} a \sin\theta+\frac{m}{\sin\theta} .
\end{align}
The general perturbation can be expressed as a superposition of different modes with a time and a $\varphi$ dependence given by $e^{i(\sigma^{\dagger} t + m \varphi)}$.
The intrinsic derivatives in Eqs. (\ref{decouple}) and (\ref{T4}) then become \begin{align}\label{derivative}
D=\mathscr{D}_{0}, \qquad &\Delta=-\frac{\Delta_{r}}{2\Sigma }\mathscr{D}_{0}^{\dagger}, \qquad\delta=\frac{1}{\sqrt{2} \overline{\rho}}\mathscr{L}_{0}^{\dagger}, \qquad\overline\delta=\frac{1}{\sqrt{2} \overline{\rho}^{*}}\mathscr{L}_{0}.
\end{align}
Taking $\psi_{4}^{B}=(\overline\rho^{*})^{-4} \phi_{4}^{B}$, and using the spin coefficients given by (\ref{spin}), we find that the decoupled equation (\ref{decouple}) can be rewritten as
\begin{align}\label{EqT}
\Big[\Delta_{r}\big(\mathscr{D}_{-1}^{\dagger}+\frac{2\overline{\rho}\overline{\rho}^*}{\Delta_{r}}F_1 \big)\big(\mathscr{D}_{0}-\frac{3}{\overline{\rho}^{*}}\big)+F_{2}\big(\mathscr{L}_{-1}-\sqrt{2}\overline\rho^{*} F_3\big)\big(\mathscr{L}_{2}^{\dagger}-\frac{3 i a \sin\theta}{\overline{\rho}^{*}}\big)+2\overline{\rho}\overline{\rho}^*F_{4}\Big]\phi_{4}^{B}={\cal T}_4,
\end{align}
with
\begin{align}
\nonumber{\cal T}_4&=4\pi G F_{4}\Big\{\mathscr{L}_{-1}\Big[\frac{\overline\rho^{*}}{3\psi_2-2\phi_{11}}\mathscr{L}_{0} \Big(\overline\rho\overline\rho^{*2}T_{nn}\Big) \Big]+\frac{\Delta_{r}^{2}}{2}\mathscr{D}_{0}^{\dagger}\Big[\frac{\overline\rho^{*}}{3\psi_2+2\phi_{11}}\mathscr{D}_{0}^{\dagger}\Big(
\overline\rho^{-1}\overline\rho^{*2}T_{\overline{m} \overline{m}}\Big) \Big]\\ &+ \frac{\Delta_{r}^2}{\sqrt{2}}\Big\{\mathscr{D}_{0}^{\dagger}\Big[
\frac{\overline\rho^{-2}\overline\rho^{*}}{\Delta_{r}(3\psi_2+2\phi_{11})}
\mathscr{L}_{-1}\Big(
\overline\rho^{2}\overline\rho^{*2}T_{\overline{m} n}\Big) \Big]+\mathscr{L}_{-1}\Big[
\frac{\overline\rho^{-2}\overline\rho^{*}}{3\psi_2-2\phi_{11}}
\mathscr{D}_{0}^{\dagger}\Big(
\frac{\overline\rho^{2}\overline\rho^{*2}}{\Delta_{r}}T_{ \overline{m} n}\Big) \Big]\Big\}\Big\},
\label{T4TT}
\end{align}
where $T_{n n}=\phi^B_{22}/(4\pi G), $ $ T_{n \overline{m} }= \phi^B_{21}/(4\pi G), $  and  $T_{\overline{m} \overline{m} }=\phi^B_{20}/(4\pi G).$

It should be noted that although $\psi_{4}^{B}$ is decoupled by taking the gauge in which  $\psi_{1}^{B}$ and $\psi_{3}^{B}$ vanish, Eq. (\ref{EqT}) with $a_i=0$ ($i\ge 2$) is the same as that of the Kerr spacetime obtained in Ref. \cite{Teukolsky}. This clearly shows that $\psi_{4}^{B}$ is gauge invariant for the linear perturbation described by (\ref{Pmetric}).

\subsection{Radial equation and solution  of  $\psi^B_{4}$ for the gravitational perturbation in the effective rotating spacetime}

To find the radial equation and the solution of $\psi^B_{4}$,  we should separate the variables of Eq. (\ref{EqT}) and work out the tetrad components of the energy momentum tensor of the system.

\subsubsection{Separation of variables of $\psi^B_{4}$ in the slowly rotating case }

It is clear that Eq. (\ref{EqT}) does not appear to be separable directly.
However, if a slowly rotating background is close enough to spherical symmetry, we can separate the perturbation gravitational wave equation (\ref{EqT}) in the radial and angular parts. In principle, the method can be extended to any order of the rotational parameter. Here, the perturbation equation is derived explicitly up to the first order.

For the slowly rotating case, we assume that $\phi_{4}^{B}$ can be decomposed into Fourier harmonic components according to
\begin{equation}
\phi_{4}^{B}=\sum_{ \ell m} \frac{1}{\sqrt{2\pi} } \int d\omega e^{-i(\omega t-m\varphi )}\,_{-2} Y_{\ell m}(\theta ) R_{\ell m}(r),
\end{equation}
where we take $\sigma^{\dagger} =-\omega$, the angular function$\ _{-2} Y_{\ell m}(\theta )$ is called the spin-weighted spheroidal harmonic, which is normalized as
\begin{equation}\label{nor}
\int_{0}^{\pi}\,_{-2}Y_{\ell m}^{*}(\theta) \,_{-2}\,Y_{\ell m}(\theta) \sin\theta \mathrm{d}\theta =1.
\end{equation}
We can then expand Eq. (\ref{EqT}) to the first order in the rotation parameter $a$, which is expressed as
\begin{align}
{\cal A}(\ell) \, \phi_{4}^{B} + a \, {\cal B}\, \phi_{4}^{B}- a\, \cos\theta \, {\cal C}(\ell)\, \phi_{4}^{B}= {\cal T}^{(0)}_{4} +a {\cal T}^{(1)}_{4} ,\label{Eqa}
\end{align}
with
\begin{align}\label{abc}
&{\cal A}(\ell)=\Delta^0_{r}\Big[\mathscr{D}_{-1}^{0 \dagger} \mathscr{D}^0_{0}-\big(\frac{3}{r} \mathscr{D}_{-1}^{0 \dagger}+ F_{10} \mathscr{D}^0_{0} \big)+\frac{3}{r^2}\big(1+r F_{10}\big)\Big]+2 r^2 F_{40}-F_{20} \lambda(\ell),\nonumber \\
&{\cal B}= i m\Big[\Big(\frac{3}{r}-F_{10}-\frac{2 \Delta^{0'}_{r}}{\Delta^0_{r}}+ \frac{2 i\,r^2 \omega}{\Delta^0_{r}}\Big)-2 F_{20}\big( \,i\,\omega+\frac{3}{r} \big)\Big] ,\nonumber \\
&{\cal C}(\ell)=i \Delta^0_{r}\Big[\big(\frac{3}{r^2} \mathscr{D}_{-1}^{0 \dagger}+ F_{11} \mathscr{D}^0_{0} \big)-\frac{3}{r^3}\big(2+r F_{10}+r^2 F_{11}\big)\Big]\nonumber \\ &\ \ \ \ \ \ \ \ +2 F_{20}\big( \omega-\frac{3 i}{r} \big)-2 i r^2 F_{41} -i F_{21}\,\lambda(\ell),\nonumber \\
&\lambda(\ell)=(\ell+2)(\ell-1),
\end{align}
where we use the relation $(\mathscr{L}^0_{-1} \mathscr{L}^{0\dag}_{2})\,_{-2}Y_{\ell m}(\theta)=-\lambda(\ell)\,_{-2}Y_{\ell m}(\theta)$. Here and hereafter all quantities with the superscripts ``$\,^0$" mean that they are fixed at $a=0$, and the functions $F_{ij}$ ($i=1,2,3,4$, and $j=0,1$) are defined as
\begin{align}
&F_{10}=-\frac{32 \Delta^0_r-23 r \Delta^{0'}_r+7 r^2 \Delta^{0''}_r-r^3 \Delta^{0(3)}_r}{r\big[8 \Delta^0_r-r( 5 \Delta^{0'}_r-r \Delta^{0''}_r)\big]} ,\nonumber \\
&F_{11}=-\frac{3}{r^2\big[8 \Delta^0_r-r( 5 \Delta^{0'}_r-r \Delta^{0''}_r)\big]^2} \Big\{32(\Delta^{0}_r)^2+4 r \Delta^{0}_r \Big[r\big(4+r \Delta^{0(3)}\big)-10\Delta^{0'}_r\Big]\nonumber\\
&+r^2\Big[17 (\Delta^{0'}_r)^2+r^2\Big(\Delta^{0''}_r\big(8+\Delta^{0''}_r\big)-2 r \Delta^{0(3)}_r\Big)-r\Delta^{0'}_r\Big(16+6\Delta^{0''}_r+r\Delta^{0(3)}_r\Big)\Big]\Big\} ,\nonumber \\
&F_{20}=\frac{8 \Delta^{0}_r-r( 5 \Delta^{0'}_r-r \Delta^{0''}_r)}{4 \Delta^{0}_r-r( 2r + \Delta^{0'}_r)},\nonumber \\
&F_{21}=-\frac{3}{r}\,\frac{4 \Delta^0_r-r\big[ 4 \Delta^{0'}_r-r (2+ \Delta^{0''}_r)\big]}{4 \Delta^0_r-r( 2r + \Delta^{0'}_r)},\nonumber \\
&F_{40}=\frac{1}{2 r^4}\,\big[8 \Delta^0_r-r( 5 \Delta^{0'}_r-r \Delta^{0''}_r)\big],\nonumber \\
&F_{41}= \frac{3}{2 r^5}\,\big[4 \Delta^0_r-r( 2r + \Delta^{0'}_r)\big] .
\end{align}
By means of the identity \cite{Kojima}
\begin{align}
\cos \theta{}_{s} Y_{\ell m}&=\mathcal{Q}_{\ell+1}{}_{s} Y_{\ell+1 m}-\frac{m s}{\ell \left(\ell +1\right)}{}_{s}Y_{\ell m}+\mathcal{Q}_{\ell}{}_{s} Y_{\ell-1 m}, 
\end{align}
with $\mathcal{Q}_{\ell}=\frac{1}{\ell} \sqrt{\frac{(\ell-s)(\ell+s)(\ell-m)(\ell+m)}{(2\ell-1)(2\ell+1)}}$, we find that the separation of variables of the gravitational wave equation Eq. (\ref{Eqa}) in the radial parts for the slowly rotating case is explicitly given by
\begin{align}
{\cal A}(\ell)& \, R_{\ell m } + a \Big[ {\bar{\cal B}}\, R_{\ell m }- {\cal C}(\ell-1,m) {\cal Q}_{\ell} \, R_{\ell-1 m }- {\cal C}(\ell+1,m) {\cal Q}_{\ell+1} \, R_{\ell+1 m }\Big]= T_{\ell m \omega}(r),\label{EqaT}
\end{align}
where $\bar{\cal B}={\cal B}+\frac{m s}{\ell \left(\ell +1\right)} {\cal{C}}(\ell,m) $, and 
\begin{eqnarray}
&&T_{\ell m \omega}(r)=\frac{1 }{2\pi } \int_{-\infty }^{+\infty} dt \int d\Omega \Big({\cal T}^{(0)}_{4} +a {\cal T}^{(1)}_{4} \Big) \ e^{i(\omega t-m \varphi)} \frac{ \ _{-2}Y^*_{\ell m}(\theta ) }{\sqrt{2\pi} }.\label{VrgenT}
\end{eqnarray}

\subsubsection{Tetrad components of the energy-momentum tensor }

By means of the definition in Ref. \cite{Sasaki81}, we know that the energy-momentum tensor for the EOB theory, i.e., a particle that orbits around a massive black hole described by the effective metric, can be expressed as
\begin{equation}
T^{\mu\nu}
=\frac{m_0}{ \Sigma \sin\theta dt/d\tau}\frac{d x^{\mu}}{ d\tau}
\frac{d x^{\nu}}{ d\tau}\delta(r-r(t))\delta(\theta-\theta(t))
 \delta(\varphi-\varphi(t)),
\end{equation}
where $x^\mu=\bigl(t,r(t),\theta(t),\varphi(t)\bigr)$ is a geodesic trajectory and $\tau$ is the proper time along the geodesic.
The geodesic equations in the effective metric are
\begin{eqnarray}
& &\Sigma \frac{d\theta }{ d\tau}
      =\pm\Bigl[{\mathscr C}-\cos^2\theta \Bigl\{a^2(1-E^2)+
      \frac{L^2 }{ \sin^2\theta}\Bigr\}\Bigr]^{1/2}
   ,\nonumber\\
& &\Sigma\frac{d\varphi }{ d \tau}
=-\Big(a E-\frac{L }{ \sin^2\theta}\Big)+\frac{a }{\Delta_r}\Big(E (r^2+a^2)-a L\Big),\nonumber\\
& &\Sigma\frac{dt }{d\tau}=
      -\Bigl(aE-\frac{L }{ \sin^2\theta}\Bigr)a\sin^2\theta
      +\frac{r^2+a^2 }{ \Delta_r}\Bigl(E(r^2+a^2)-a L\Bigr)
     , \nonumber\\
& &\Sigma\frac{dr }{ d\tau}=\pm\Big[[E(r^2+a^2)-a L]^2-\Delta_r [(Ea-L)^2+r^2+{\mathscr C}]\Big]^{1/2},
\label{geodesicgen}
\end{eqnarray}
where $E$, $L$, and ${\mathscr C}$ are the energy, the $z$-component of the angular momentum \cite{Chen}, and the Carter constant of a test particle,
respectively.

The tetrad components of the energy-momentum tensor can then be expressed as
\begin{eqnarray}
T_{n n}&=&m_0\frac{C_{n n} }{ \sin\theta}
\delta(r-r(t)) \delta(\theta-\theta(t)) \delta(\varphi-\varphi(t)),
\nonumber\\
T_{{\overline m} n}&=&m_0\frac{C_{{\overline m} n} }{ \sin\theta}
\delta(r-r(t)) \delta(\theta-\theta(t)) \delta(\varphi-\varphi(t)),\nonumber\\
T_{{\overline m} {\overline m}}&=&m_0
\frac{C_{{\overline m} {\overline m}} }{ \sin\theta}
\delta(r-r(t)) \delta(\theta-\theta(t)) \delta(\varphi-
\varphi(t)),\label{tij}
\end{eqnarray}
with
\begin{eqnarray}
&&C_{n n}=\frac{1}{ 4\Sigma^3 \dot t}\left[E(r^2+a^2)-a L
+\Sigma\frac{dr}{ d\tau} \right]^2,\nonumber\\
&&C_{{\overline m} n}=
-\frac{1 }{ 2\sqrt{2}\overline{\rho}^*\Sigma^2 \dot t}\left[E(r^2+a^2)-a L
+\Sigma\frac{dr}{ d\tau} \right]
\left[i\sin\theta\Bigl(aE-\frac{L }{ \sin^2\theta}\Bigr)
+\Sigma \frac{d\theta}{d\tau}\right],
\nonumber\\
&&C_{{\overline m} {\overline m}}=
\frac{1 }{ 2\overline{\rho}^{*2} \Sigma \dot t }
\left[i\sin\theta
\Bigl(aE-\frac{L }{ \sin^2\theta}\Bigr)
+\Sigma \frac{d\theta}{d\tau}\right]^2,
\label{cij}
\end{eqnarray}
where $\dot t=dt/d\tau$.

Using Eqs.~(\ref{T4TT}), (\ref{VrgenT}), and (\ref{tij}), and performing integration by part, we obtain
\begin{eqnarray}
&&T_{\ell m \omega}(r)
=
-\frac{4m_0 G}{\sqrt{2\pi}}\int^{\infty}_{-\infty}
dt\int d\theta e^{i\omega t-im\varphi(t)} \nonumber\\
&&
\times\Bigg\{-\frac{1}{ 2}\mathscr{L}_1^{\dag} \Big[ \frac{\overline\rho^{*} }{F_5}\mathscr{L}_2^{\dag}( \ _{-2}Y_{\ell m} F_4)
\Big]
C_{n n}\overline\rho^{*2} \overline\rho \delta(r-r(t))
\delta(\theta-\theta(t)) \nonumber\\
&&
+\frac{\Delta_{r}^2 \overline\rho^{*} }{ 2 \sqrt{2} \overline\rho^{2}}
\Big((1+F_2) \mathscr{L}_2^{\dag} \ _{-2}Y_{\ell m}+\frac{ \ _{-2}Y_{\ell m} \overline\rho^{2}}{\overline\rho^{*}} \frac{\partial }{\partial \theta}\frac{ \overline\rho^{*}}{\overline\rho^{2}}+\frac{ \ _{-2}Y_{\ell m}}{F_5}\frac{\partial F_4}{\partial \theta}\Big)
\nonumber \\ && \times \mathscr{D}_{0}^{\dagger} \Big[
\frac{C_{{\overline m} n} \overline\rho^{*2} \overline\rho^{2}}{\Delta_{r}}
\delta(r-r(t))\delta(\theta-\theta(t)) \Big] \nonumber\\
&&
+\frac{1}{ 2\sqrt{2} }
\mathscr{L}_2^{\dag}\Big[ F_4 \ _{-2}Y_{\ell m}\frac{\partial }{\partial r}\Big(\frac{\overline\rho^{*}}{\overline\rho^{2} F_4}\Big) \Big]
C_{{\overline m} n}\Delta_{r} \ \overline\rho^{*2}\overline\rho^{2}
\delta(r-r(t))\delta(\theta-\theta(t)) \nonumber\\
&&
-\frac{1}{ 4} \Delta_{r}^2 \ _{-2}Y_{\ell m} F_4 \mathscr{D}_{0}^{\dagger}\Big[\frac{\overline\rho^{*}}{F_4}
\mathscr{D}_{0}^{\dagger} \Big(\overline\rho^{-1} \overline\rho^{*2} C_{{\overline m}{\overline m}}
\delta(r-r(t))\delta(\theta-\theta(t))\Big) \Big]
\Bigg\}, 
\label{Tgen}
\end{eqnarray}
where $F_5=3\psi_4^B-2 \phi_{11}$. We now expand Eq. (\ref{Tgen}) for $ T_{\ell m \omega}(r)$ to the first order in the rotation parameter $a$, i.e.,
\begin{eqnarray} \label{expandT}
 T_{\ell m \omega}(r)= T^{(0)}_{\ell m \omega}(r)+a\, T^{(1)}_{\ell m \omega}(r),
\end{eqnarray}
with
\begin{eqnarray}
&&T^{(0)}_{\ell m \omega}(r)=\frac{1 }{2\pi } \int_{-\infty }^{+\infty} dt \int d\Omega \ {\cal T}^{(0)}_{4} \ e^{i(\omega t-m \varphi)} \frac{ \ _{-2}Y^*_{\ell m}(\theta ) }{\sqrt{2\pi} },\label{VrgenTT0}\\
&&T^{(1)}_{\ell m \omega}(r)=\frac{1 }{2\pi } \int_{-\infty }^{+\infty} dt \int d\Omega \ {\cal T}^{(1)}_{4} \ e^{i(\omega t-m \varphi)} \frac{ \ _{-2}Y^*_{\ell m}(\theta ) }{\sqrt{2\pi} }.\label{VrgenTT1}
\end{eqnarray}

\vspace{0.5cm}

 {\bf 2-1. Zeroth-order source}

For a source bounded in a finite range of $r$, it is convenient to rewrite the zeroth order of the source Eq.~(\ref{VrgenTT0}) as
\begin{eqnarray}
T^{(0)}_{\ell m\omega}(r)&=&-m_0 G \int^{\infty}_{-\infty}dt
e^{i\omega t-i m \varphi(t)}
(\Delta^0_r )^2\Big\{\big(A^{(0)}_{nn\,0}+A^{(0)}_{{\overline m}n\,0}+
A^{(0)}_{{\overline m}{\overline m}\,0}\big)\delta(r-r(t)) \nonumber\\
&& +\Big[\big(A^{(0)}_{{\overline m}n\,1}+A^{(0)}_{{\overline m}{\overline m}\,1}\big)
\delta(r-r(t))\Big]'
+\Big[A^{(0)}_{{\overline m}{\overline m}\,2}
\delta(r-r(t))\Big]''\Big\},
\label{TgenTTsl}
\end{eqnarray}
with
\begin{eqnarray}
A^{(0)}_{nn\,0}&=&-\frac{2\,r^4 }{ \sqrt{2\pi}\,(\Delta^0_r )^2}\,
C^{(0)}_{n n} \,F_{20} \,
\mathscr{L}_1^{0\dag}\Big[\mathscr{L}_2^{0\dag}\Big(\,_{-2}Y_{\ell m}(\theta)\Big) \Big],\nonumber \\
A^{(0)}_{{\overline m}n\,0}&=&\frac{r^3 }{ \sqrt{\pi}\Delta^0_r }\, C^{(0)}_{{\overline m} n} \Big[-\big(1+F_{20}\big)\frac{i K^0 }{ \Delta^0_r }+\frac{F_{20}}{r} -F_{20,r}-\frac{F_{40,r}}{F_{40}}\Big] \mathscr{L}_2^{0\dag} \Big(\ _{-2}Y_{\ell m}(\theta) \Big),\nonumber \\
A^{(0)}_{{\overline m}{\overline m}\,0}
&=&-\frac{r^2 }{ \sqrt{2\pi}}\,
C^{(0)}_{{\overline m} {\overline m}}\ _{-2}Y_{\ell m}(\theta) \Bigl[
i\Bigl(\frac{K^0 }{ \Delta^0_r }\Bigr)' -\frac{(K^0)^2 }{ (\Delta^0_r )^2}\nonumber \\ &&+ \frac{i K^0 }{ \Delta^0_r }\Big(\frac{1}{r}+\frac{F_{40,r}}{F_{40}}\Big)+\frac{F_{40,r}}{r F_{40}}+\Big(\frac{F_{40,r}}{F_{40}}\Big)' \Bigr],\nonumber \\
A^{(0)}_{{\overline m}n\,1}&=&\frac{
 r^3}{ \sqrt{\pi}\Delta^0_r}\,C^{(0)}_{{\overline m} n}
\big(1+F_{10}\big) \mathscr{L}_2^{0\dag}\Big(\ _{-2}Y_{\ell m}(\theta) \Big)
,\nonumber \\
A^{(0)}_{{\overline m}{\overline m}\,1}
&=&\frac{ r^2 }{ \sqrt{2\pi}}
\,
C^{(0)}_{{\overline m}{\overline m}}\ _{-2}Y_{\ell m}(\theta)
\Bigl( 2 i\frac{K^0 }{ \Delta^0_r }+\frac{1}{r}+\frac{F_{40,r}}{F_{40}}\Bigr),\nonumber \\
A^{(0)}_{{\overline m}{\overline m}\,2}
&=&-\frac{r^2}{ \sqrt{2\pi}}\,
C^{(0)}_{{\overline m} {\overline m}}\ _{-2}Y_{\ell m}(\theta) ,
\label{Aijs}
\end{eqnarray}
where the functions $C^{(0)}_{ij}$ ($i,j=n,\overline m$) are given by
\begin{eqnarray}
C^{(0)}_{n n}&=&\frac{\Delta^0_{r}}{ 4 r^4 E}\left(E
+ \frac{dr}{ d\tau} \right)^2,\nonumber\\
C^{(0)}_{{\overline m} n}&=&
\frac{\Delta^0_{r} }{ 2\sqrt{2} r^3 E}\left(E
+\frac{dr}{ d\tau} \right)
\left(\frac{ i L }{ r^2\sin\theta}
- \frac{d\theta}{d\tau}\right),
\nonumber\\
C^{(0)}_{{\overline m} {\overline m}}&=&
\frac{\Delta^0_{r} }{ 2 r^2 E }
\left( \frac{d\theta}{d\tau}-\frac{ i L }{ r^2 \sin\theta}
\right)^2.
\label{cijslS}
\end{eqnarray}

\vspace{0.5cm}
 {\bf 2-2. First-order source}

The first order of the source Eq.~(\ref{VrgenTT1}) can be expressed as
\begin{eqnarray}
T^{(1)}_{\ell m\omega}(r)&=&-m_0 G \int^{\infty}_{-\infty}dt
e^{i\omega t-i m \varphi(t)}
(\Delta^0_r )^2\Big\{\big(A^{(1)}_{nn\,0}+A^{(1)}_{{\overline m}n\,0}+
A^{(1)}_{{\overline m}{\overline m}\,0}\big)\delta(r-r(t)) \nonumber\\
&&+\Big[\big(A^{(1)}_{{\overline m}n\,1}+A^{(1)}_{{\overline m}{\overline m}\,1}\big)
\delta(r-r(t))\Big]'
+\Big[A^{(1)}_{{\overline m}{\overline m}\,2}
\delta(r-r(t))\Big]''\Big\},
\label{TgenTTsl1}
\end{eqnarray}
with
\begin{align}
&A^{(1)}_{nn\,0}=-\frac{2 r^4 \,C^{(0)}_{nn}}{ \sqrt{2\pi}\Delta_r^{0\,2}}\Bigg\{ F_{20} \Big[\Big(\omega\, \sin\theta +\big(\frac{C^{(1)}_{nn}}{\,C^{(0)}_{nn}}-\frac{i\, \cos\theta}{r}\big)\mathscr{L}_1^{0+}\Big)\mathscr{L}_2^{0+}+\omega \mathscr{L}_1^{0+} \sin\theta\Big]\,_{-2}Y_{\ell m}(\theta)\nonumber \\
&\ \ \ \ \ \ \ -\frac{i\,F_{40}}{F_{50}}\Big[ \Big(\frac{F_{51}}{F_{50}}+\frac{1}{r}\big) \mathscr{L}_1^{0+}\cos\theta\mathscr{L}_2^{0+} \big(\,_{-2}Y_{\ell m}(\theta)\big)-\frac{F_{41}}{F_{40}}
\mathscr{L}_1^{0+}\mathscr{L}_2^{0+} \big( \cos\theta \,_{-2}Y_{\ell m}(\theta)\big)\Big]\Bigg\}
,\nonumber \\
&A^{(1)}_{{\overline m}n\,0}=-\frac{r^4}{ \sqrt{\pi}\Delta^0_r }\Bigg\{\frac{ 1 }{r^3}\Big[ \Big(1+r \frac{F_{40}'}{F_{40}}\Big) r C^{(1)}_{{\overline m} n} \mathscr{L}_2^{0\dag}-i \Big(6+3 r \frac{F_{40}'}{F_{40}}-r^2 \frac{F_{41}'}{F_{40}}\big) C^{(0)}_{{\overline m} n} \mathscr{L}_2^{0\dag} \cos\theta \nonumber \\ &\ \ \ \ \ \ \ - i r^2 \frac{F_{40}'}{F_{40}} \frac{F_{41}}{F_{40}} C^{(0)}_{{\overline m} n}\mathscr{L}_2^{0\dag} \cos\theta \Big] +\frac{1}{r^2}\big(1+r \frac{ F_{40}' }{ F_{40}}\big)C^{(0)}_{{\overline m} n}\, \omega \,\sin\theta \nonumber \\ &\ \ \ \ \ \ \ + \Big\{\big(C^{(1)}_{{\overline m} n} \mathscr{D}^0_{0}+\frac{ i\,m}{\Delta^0_r}\,C^{(0)}_{{\overline m} n}\big)\frac{1+F_{20}}{r} -3 i\,\cos\theta \mathscr{D}^0_{0} \frac{1+F_{20}}{r^2}\Big] \Big\}\mathscr{L}_2^{0\dag}\nonumber \\ &\ \ \ \ \ \ \ + C^{(0)}_{{\overline m} n}\mathscr{D}^0_{0} \Big[ \frac{1}{r}\Big(i\, F_{21}\, \cos\theta \,\mathscr{L}_2^{0\dag}+\Big((1+F_{20})\omega+\frac{3\, i}{r}-\frac{i\, F_{41}}{F_{50}}\Big)\sin\theta\Big)\Big]\Bigg\} \,_{-2}Y_{\ell m}(\theta),\nonumber \\
&A^{(1)}_{{\overline m}{\overline m}\,0}
=-\frac{r^2 }{ \sqrt{2\pi}}\Big( C^{(1)}_{{\overline m} {\overline m}} \mathscr{H}_0+C^{(0)}_{{\overline m} {\overline m}} \mathscr{H}_1-\frac{4\, i\, a\, \cos\theta }{r}\,C^{(0)}_{{\overline m} {\overline m}} \mathscr{H}_0\Big)\,_{-2}Y_{\ell m}(\theta),\nonumber \\
&A^{(1)}_{{\overline m}n\,1}=\frac{r^3}{ \sqrt{\pi}\Delta^0_r }\Bigg\{\big(1+F_{20}\big )\Big( C^{(1)}_{{\overline m} n}-\frac{3\, i\, \cos\theta}{r} C^{(0)}_{{\overline m} n}\Big)\mathscr{L}_2^{0\dag} \nonumber \\ &\ \ \ \ \ \ \ + C^{(0)}_{{\overline m} n}\Big[ i\, F_{21}\, \cos\theta \,\mathscr{L}_2^{0\dag}+\Big((1+F_{20})\omega+\frac{3\, i}{r}+\frac{i\, F_{41}}{F_{50}}\Big)\sin\theta\Big]\Bigg\} \,_{-2}Y_{\ell m}(\theta),\nonumber \\
&A^{(1)}_{{\overline m}{\overline m}\,1}
=\frac{r^2}{ \sqrt{2\pi}}\Bigg\{ C^{(0)}_{{\overline m} {\overline m}}\Big(i\, \cos\theta \,\frac{F_{40}\,F_{41}'-F_{41}\,F_{40}'}{F_{40}^2}+\frac{2 \,i\, m}{\Delta^0_r}+\frac{i \,\cos\theta}{r^2}\Big) \nonumber \\ &\ \ \ \ \ \ \ \ +\Bigl( 2\, i\,\frac{K^0 }{ \Delta^0_r }+\frac{1}{r}+\frac{F_{40,r}}{F_{40}}\Bigr)\Big( C^{(1)}_{{\overline m} {\overline m}}-\frac{4\, i\, \cos\theta}{r} C^{(0)}_{{\overline m} {\overline m}}\Big)\Bigg\}\,_{-2}Y_{\ell m}(\theta),\nonumber \\
&A^{(1)}_{{\overline m}{\overline m}\,2}
=\frac{r^2}{ \sqrt{2\pi}}\left[
\frac{4 i \cos\theta}{r}C^{(0)}_{{\overline m} {\overline m}}-C^{(1)}_{{\overline m} {\overline m}}\right]\,_{-2}Y_{\ell m}(\theta),
\end{align}
where
\begin{eqnarray}
C^{(1)}_{n n}&=&\frac{L}{ r^4 }\frac{r^2-\Delta^0_r}{E} C^{(0)}_{n n}-\frac{L \Delta^0_r (E
+ dr/ d\tau)}{2 E r^6} ,\nonumber\\
C^{(1)}_{{\overline m} n}&=&\left(\frac{L}{ r^4 }\frac{r^2-\Delta^0_r}{E} +\frac{i \cos\theta}{r}\right) C^{(0)}_{{\overline m} n}-\frac{\Delta^0_r }{ 2\sqrt{2} r^5 E}\left[i E \sin\theta\big(E
+\frac{dr}{ d\tau} \big)+L
\big(\frac{ i L }{r^2 \sin\theta}
- \frac{d\theta}{d\tau}\big)\right],
\nonumber\\
C^{(1)}_{{\overline m} {\overline m}}&=&\left(\frac{L}{ r^4 }\frac{r^2-\Delta^0_r}{E} +\frac{2 i \cos\theta}{r}\right)C^{(0)}_{{\overline m} {\overline m}}+\frac{i \sin\theta \Delta^0_r }{ r^4 }
\left( \frac{d\theta}{d\tau}-\frac{ i L }{r^2 \sin\theta}
\right),\nonumber \\
\mathscr{H}_0&=&
i\Bigl(\frac{K^0 }{ \Delta^0_r}\Bigr)' -\frac{(K^0)^2 }{ (\Delta^0_r )^2}+ \frac{i K^0 }{ \Delta^0_r }\Big(\frac{1}{r}+\frac{F_{40,r}}{F_{40}}\Big)+\frac{F_{40,r}}{r F_{40}}+\Big(\frac{F_{40,r}}{F_{40}}\Big)' , \nonumber \\
\mathscr{H}_1&=&\frac{m}{\Delta^0_r} \Big[i \Big(\frac{1}{r}+\frac{F_{40}'}{F_{40}}-\frac{\Delta^{0'}_r}{\Delta^0_r}\Big)-\frac{2 K^0}{\Delta^0_r}\Big] -\frac{K^0\,\cos\theta}{\Delta^0_r} \Big[\frac{1}{r^2}+\frac{1}{F_{40}^2}\Big(F_{40}F_{41}'-F_{41}F_{40}'\Big)\Big]\nonumber \\
&+&\frac{i\,\cos\theta}{r^2\,F_{40}^2} \Big(F_{40}F_{40}'-r\,F_{41}F_{40}'+r\,F_{40}F_{41}'\Big)-\frac{2\,i\,\cos\theta}{F_{40}^3} \Big(F_{40}F_{40}'F_{41}'-F_{41}F_{40}^{'2}\Big)\nonumber \\
&+&\frac{i\,\cos\theta}{F_{40}^2} \Big(F_{40}F_{41}''-F_{41}F_{40}^{''}\Big).
\label{cijslS1}
\end{eqnarray}

\subsubsection{Solution of $\psi^B_{4}$ in the effective rotating spacetime}

Using Eq. (\ref{expandT}), we can rewrite Eq. (\ref{EqaT}) as
\begin{align}
{\cal A}(\ell)& \, R_{\ell m } + a \Big[ {\bar{\cal B}}\, R_{\ell m }- {\cal C}(\ell-1,m) {\cal Q}_{\ell} \, R_{\ell-1 m }- {\cal C}(\ell+1,m) {\cal Q}_{\ell+1} \, R_{\ell+1 m }\Big]\nonumber \\
&= T^{(0)}_{\ell m \omega}(r)+a\, T^{(1)}_{\ell m \omega}(r). \label{EqaTS}
\end{align}
It is clear that Eq. (\ref{EqaTS}) contains three modes, i.e., $R_{\ell m }$, $R_{\ell-1 m }$, and $R_{\ell+1 m }$. Therefore, for a calculation that is accurate to the first order in $a$, the $ R_{\ell m }$ and $ R_{\ell\pm1 m }$ terms in the square brackets should be evaluated at the lower order. That is, we can solve Eq. (\ref{EqaTS}) order by order by expanding the function $ R_{\ell m }$ to the first order in the rotational parameter $a$ as
\begin{align}
R_{\ell m }=R^{(0)}_{\ell m }+a \,R^{(1)}_{\ell m }. \label{EqaR}
\end{align}
The equations for the zeroth and first orders of the rotational parameter $a$ are respectively given by
\begin{eqnarray}
&&{\cal A}(\ell) \, R^{(0)}_{\ell m }= T^{(0)}_{\ell m \omega}(r), \label{EqaTT0} \\ \label{EqaTT1}
&&{\cal A}(\ell) \, R^{(1)}_{\ell m } + \Big[ {\bar{\cal B}}\, R^{(0)}_{\ell m }- {\cal C}(\ell-1) {\cal Q}_{\ell} \, R^{(0)}_{\ell-1 m }- {\cal C}(\ell+1) {\cal Q}_{\ell+1} \, R^{(0)}_{\ell+1 m }\Big]
=T^{(1)}_{\ell m \omega}(r).
\end{eqnarray}

\vspace{0.5cm}

 {\bf 3-1. Formal solution of the zeroth-order equation}

The zeroth-order equation can be obtained by substituting Eq. (\ref{abc}) into Eq. (\ref{EqaTT0}), which can be expressed explicitly as
\begin{eqnarray} \label{EqTS0}
\left[r^{3}F_{40} (\Delta^0_{r})^{2} \frac{\mathrm{d} }{\mathrm{d} r} \Big(\frac{1}{r^{3} F_{40} \Delta^0_{r}} \frac{\mathrm{d} }{\mathrm{d} r} \Big) + V \right] R^{(0)}_{\ell m\omega}(r)= T^{(0)}_{\ell m \omega}(r),
\end{eqnarray}
where
\begin{align} \label{VrS0}
&V= \frac{r^{2} \omega(r^{2} \omega+2 i \Delta^{0'}_{r}) }{\Delta^0_{r}} -i r \omega \Big(5 - r F_{10} \Big)+\frac{3 [\Delta^0_{r}(1+r F_{10})+ r \Delta^{0'}_{r}]}{r^{2}}+2 r^2 F_{40}-\lambda(\ell) F_{20}.\nonumber
\end{align}
It is interesting to note that if we take $a_i=0$ ($i\ge 2$), Eq. (\ref{EqTS0}) will reduce to the equation of the gravitational wave for $\psi^B_4$ in the Schwarzschild spacetime \cite{Teukolsky}.

The asymptotical homogeneous solutions for Eq.~\eqref{EqTS0} are \begin{eqnarray}
R^{\rm in(0)}_{asy}
\to \left\{\begin{array}{cc}\ \
B^{\rm trans}_{\ell m\omega}(\Delta_r^0)^2 e^{-i \omega r^*},
\ \ \ \ \ \ \ \ \ \ \ \ \ \ \ \ \ \ \ \ \ \ \ \ \ & \text{for} \ \ r\to r_+, \ \ \ \\
r^3 B^{\rm ref}_{\ell m\omega}e^{i\omega r^*}+
r^{-1}B^{\rm inc}_{\ell m\omega} e^{-i\omega r^*},
\ \ \ \ \ \ & \text{for} \ \ r\to +\infty,
\end{array}\right.
\label{Kk}
\end{eqnarray}

\begin{eqnarray}
R^{\rm up(0)}_{asy}
\to \left\{\begin{array}{cc}\ \
C^{\rm up}_{\ell m\omega} e^{i \omega r^*}+
(\Delta_r^0)^2 C^{\rm ref}_{\ell m\omega} e^{-i\omega r^*},
\ \ \ \ \ & \text{for} \ \ r\to r_+, \ \ \ \ \\
C^{\rm trans}_{\ell m\omega} r^3 e^{i\omega r^*},
\ \ \ \ \ \ \ \ \ \ \ \ \ \ \ \ \ \ \ \ \ \ \ & \text{for} \ \ r\to +\infty,
\end{array}\right.
\label{Kkk}
\end{eqnarray}
where $ r^{*}=\int \frac{r^2}{ \Delta_r^0}dr $.
Moreover, the inhomogeneous solution for Eq.~\eqref{EqTS0} is
\begin{eqnarray}\label{EqTS0S}
R^{(0)}_{\ell m\omega}(r)&=&\frac{1}{2 i \omega C^{\rm trans}_{\ell m\omega}
     B^{\rm inc}_{\ell m\omega}}
 \Bigg\{R^{\rm up (0)}_{\ell m\omega}(r)\int^r_{r_+}d\tilde{r}
\frac{ R^{\rm in (0)}_{\ell m\omega}(\tilde{r} )
T^{(0)}_{\ell m\omega}(\tilde{r} ) }{\tilde{r} ^3 F_{40}(\tilde{r} ) (\Delta_r^0)^{2} (\tilde{r} ) }\nonumber \\ &+& R^{\rm in(0)}_{\ell m\omega} (r) \int^\infty_{r}d\tilde{r}
\frac{R^{\rm up (0)}_{\ell m\omega}(\tilde{r} ) T^{(0)}_{\ell m\omega}(\tilde{r} ) }{\tilde{r} ^3 F_{40}(\tilde{r} ) (\Delta_r^0)^{2}(\tilde{r} ) }\Bigg\},
\end{eqnarray}
where $R^{\rm up (0)}_{\ell m\omega}(\tilde{r} ) $ and $R^{\rm in (0)}_{\ell m\omega}(\tilde{r} ) $ are the homogeneous solutions of the radial equation \eqref{EqTS0}. Therefore, the inhomogeneous solution for Eq. (\ref{EqTS0}) at the infinity is
\begin{equation}
R^{(0)}_{\ell m\omega}(r\to\infty) \to \frac{r^3e^{i\omega r^*} }{ 2i\omega
   B^{\rm inc}_{\ell m\omega}}
\int^{\infty}_{r_+}d\tilde{r} \frac{T^{(0)}_{\ell m\omega}(\tilde{r} )
R^{\rm in (0)}_{\ell m\omega}(\tilde{r} )
}{ \tilde{r} ^3 F_{40}(\tilde{r} ) (\Delta^0_{r})^{2}}
\equiv \tilde Z^{(0)}_{\ell m\omega} r^3 e^{i\omega r^*}\,,
\label{Infinfty}
\end{equation}
with
\begin{eqnarray}
\tilde Z^{(0)} _{\ell m\omega}&=&
\frac{\pi m_0 G}{i\omega B^{\rm inc}_{\ell m\omega}}
\Biggl[ A^{(0)}_0
\frac{R^{\rm in(0)}_{\ell m\omega}(r)}{r^3 F_{40}}- A^{(0)}_1
\Big(\frac{R^{\rm in(0)}_{\ell m\omega}(r)}{r^3 F_{40}}\Big)' + A^{(0)}_2 \Big(\frac{R^{\rm in(0)}_{\ell m\omega}(r)}{r^3 F_{40}}\Big)'' \Biggr]_{r_0,\theta_0}\delta(\omega-\omega_n)\,,\nonumber \\
\label{ZZZSch}
\end{eqnarray}
where $\omega_n=m\,\Omega$, $A^{(0)}_0 =A^{(0)} _{nn\,0}+A ^{(0)}_{{\overline m}n\,0}+
A^{(0)} _{{\overline m}{\overline m}\,0},$ $
 A^{(0)}_1 =A^{(0)} _{{\overline m}n\,1}+A ^{(0)}_{{\overline m}{\overline m}\,1},$ $
 A^{(0)}_2 =A^{(0)} _{{\overline m}{\overline m}\,2}$, and $\bigl(r_0,~\theta_0\bigr)$ are the values of $\bigl(r,~\theta \bigr)$ on the geodesic trajectory.
The zeroth-order formal solution of $\psi_4^B$ is then shown by
\begin{eqnarray}
\psi^{B(0)}_4&=&\frac{1}{ r}\sum_{\ell m n}
\frac{\pi m_0 G}{i\omega_n B^{\rm inc}_{\ell m\omega_n}}
\Biggl[ A^{(0)}_0
\frac{R^{\rm in(0)}_{\ell m\omega_n}(r)}{r^3 F_{40}}- A^{(0)}_1
\Big(\frac{R^{\rm in(0)}_{\ell m\omega_n}(r)}{r^3 F_{40}}\Big)' + A^{(0)}_2 \Big(\frac{R^{\rm in(0)}_{\ell m\omega_n}(r)}{r^3 F_{40}}\Big)'' \Biggr]_{r_0,\theta_0}\nonumber \\ &\times&\frac{{}_{-2}Y_{\ell m} }{ \sqrt{2\pi}}
e^{i\omega_n(r^*-t)+im\varphi}.
\label{psi41}
\end{eqnarray}

\vspace{0.5cm}
 {\bf 3-2. Formal solution of the first-order equation}

The first-order equation (\ref{EqaTT1}) can be written explicitly as
\begin{eqnarray}
&&\Big[r^{3}F_{40} (\Delta^0_{r})^{2} \frac{\mathrm{d} }{\mathrm{d} r} \Big(\frac{1}{r^{3} F_{40} \Delta^0_{r}} \frac{\mathrm{d} }{\mathrm{d} r} \Big) + V \Big] R^{(1)}_{\ell m\omega}(r)=T^{(1)}_{\ell m \omega}(r)+ T^{(1)}_{eff}(r), \label{EqTS1}
\end{eqnarray}
with
\begin{eqnarray}
T^{(1)}_{eff}(r)&=& {\cal Q}_{\ell} \Big[\frac{i \Delta^0_{r}}{r^2}\big(3+r^2 F_{11}\big)\frac{d}{d r} +U(\ell-1)\Big]\, R^{(0)}_{\ell-1 m }(r)\nonumber \\
&+& {\cal Q}_{\ell+1} \Big[\frac{i \Delta^0_{r}}{r^2}\big(3+r^2 F_{11}\big)\frac{d}{d r} +U(\ell+1)\Big]\, R^{(0)}_{\ell+1 m }(r)- {\bar{\cal B}}\, R^{(0)}_{\ell m }(r) , \label{EqTS112}
\end{eqnarray}
where $R^{(0)}_{\ell m\omega}(r ) $ is the inhomogeneous solution \eqref{EqTS0S}, and
\begin{align}
U(\ell)&=\big(r^2 F_{11} -3\big)\omega- \frac{3i }{r^3} \Big[r\,\Delta^{0'}_{r} +\Delta^0_{r} \Big(2+r F_{10}+r^2 F_{11}\Big)\Big]\nonumber \\ & +2F_{20}\Big(\omega -\frac{3i}{r} \Big)-2 i r^2 F_{41}-i F_{21}\,\lambda(\ell).
\end{align}

Now, we solve the first-order equation (\ref{EqTS1}) by means of the Green function
\cite{Sasaki,Ref:poisson,TagoshiSasaki745}. Since the homogeneous equations \eqref{EqTS0} and \eqref{EqTS1} take the same form, the asymptotic solutions for the homogeneous solutions of Eq.~\eqref{EqTS1} are
\begin{eqnarray}
R^{\rm in(1)}_{asy}
\to \left\{\begin{array}{cc}
B^{\rm trans}_{\ell m\omega}(\Delta^0_{r})^2 e^{-i \omega r^*},
\ \ \ \ \ \ \ \ \ \ \ \ \ \ \ \ \ \ \ \ \ \ \ \ \ & \text{for} \ \ r\to r_+, \ \ \ \ \\
r^3 B^{\rm ref}_{\ell m\omega}e^{i\omega r^*}+
r^{-1}B^{\rm inc}_{\ell m\omega} e^{-i\omega r^*},
\ \ \ \ \ \ & \text{for} \ \ r\to +\infty,
\end{array}\right.
\label{KkO2}
\end{eqnarray}

\begin{eqnarray}
R^{\rm up(1)}_{asy}
\to \left\{\begin{array}{cc}
C^{\rm up}_{\ell m\omega} e^{i \omega r^*}+
(\Delta^0_{r})^2 C^{\rm ref}_{\ell m\omega} e^{-i\omega r^*},
\ \ \ \ \ & \text{for} \ \ r\to r_+, \ \ \ \ \\
C^{\rm trans}_{\ell m\omega} r^3 e^{i\omega r^*},
\ \ \ \ \ \ \ \ \ \ \ \ \ \ \ \ \ \ \ \ \ \ \ & \text{for} \ \ r\to +\infty,
\end{array}\right. \ \
\label{KkkO2}
\end{eqnarray}
and the inhomogeneous solution for the radial equation~\eqref{EqTS1} is
\begin{eqnarray}
R^{(1)}_{\ell m\omega}(r)&=&\frac{1}{2 i \omega C^{\rm trans}_{\ell m\omega}
     B^{\rm inc}_{\ell m\omega}}
 \Bigg\{R^{\rm up(0)}_{\ell m\omega} (r) \int^r_{r_+}d\tilde{r}
\frac{ R^{\rm in(0)}_{\ell m\omega}(\tilde{r} )
\big(T^{(1)}_{\ell m\omega}(\tilde{r} ) + T^{(1)}_{eff}(\tilde{r} )\big)}{\tilde{r} ^3 F_{40} (\Delta^0_{r})^{2} }\nonumber \\
&+& R^{\rm in(0)}_{\ell m\omega} (r) \int^\infty_{r}d\tilde{r}
\frac{R^{\rm up(0)}_{\ell m\omega}(\tilde{r} ) \big(T^{(1)}_{\ell m\omega}(\tilde{r} ) + T^{(1)}_{eff}(\tilde{r} )\big) }{\tilde{r} ^3 F_{40} (\Delta^0_{r})^2 }\Bigg\},
\end{eqnarray}
where $R^{\rm up (0)}_{\ell m\omega}(\tilde{r} ) $ and $R^{\rm in (0)}_{\ell m\omega}(\tilde{r} ) $ are the homogeneous solutions of the radial equation \eqref{EqTS0}. Therefore, the inhomogeneous solution of Eq. (\ref{EqTS1}) at infinity can be expressed as
\begin{equation}
R^{(1)}_{\ell m\omega}(r\to\infty)
 =\frac{r^3e^{i\omega r^*} }{ 2i\omega
   B^{\rm inc}_{\ell m\omega}}
\int^{\infty}_{r_+}d\tilde{r} \frac{
R^{\rm in(0)}_{\ell m\omega}(\tilde{r} )
\big(T^{(1)}_{\ell m\omega}(\tilde{r} ) + T^{(1)}_{eff}(\tilde{r} )\big)}{ \tilde{r} ^3 F_{40} (\Delta^0_{r})^2}
\equiv \tilde Z^{(1)}_{\ell m\omega} r^3 e^{i\omega r^*}\,.
\label{Infinfty1}
\end{equation}
Inserting Eqs.~(\ref{TgenTTsl1}) and \eqref{EqTS112} into Eq.~(\ref{Infinfty1}), we have
\begin{eqnarray}
\tilde Z ^{(1)}_{\ell m\omega}&=&
\frac{m_0 G}{2i\omega B^{\rm inc}_{\ell m\omega}}
\int^{\infty}_{-\infty}dt e^{i\omega t-i m \varphi(t)}
\Bigl[ A^{(1)}_0
\frac{R^{\rm in(0)}_{\ell m\omega}(r)}{r^3 F_{40}}
- A^{(1)}_1
\Big(\frac{R^{\rm in(0)}_{\ell m\omega}(r)}{r^3 F_{40}}\Big)' + A^{(1)}_2 \Big(\frac{R^{\rm in(0)}_{\ell m\omega}(r)}{r^3 F_{40}}\Big)'' \Bigr]_{r_0, \theta_0}\nonumber \\
&+& \frac{1}{ 2i\omega
   B^{\rm inc}_{\ell m\omega}}
\int^{\infty}_{r_+}d\tilde{r} \frac{
R^{\rm in(0)}_{\ell m\omega}(\tilde{r} )}{\tilde{r} ^3 F_{40} (\Delta^0_{r})^2}\Big\{ {\cal Q}_{\ell} \Big[\frac{i \Delta^0_{\tilde{r}}}{\tilde{r}^2}\big(3+\tilde{r}^2 F_{11}(\tilde{r})\big)\frac{d}{d \tilde{r}} +U(\ell-1)\Big]\, R^{(0)}_{\ell-1 m\omega}(\tilde{r})\nonumber \\
&+& {\cal Q}_{\ell+1} \Big[\frac{i \Delta^0_{\tilde{r}}}{\tilde{r}^2}\big(3+r^2 F_{11}(\tilde{r})\big)\frac{d}{d \tilde{r}} +U(\ell+1)\Big]\, R^{(0)}_{\ell+1 m\omega}(\tilde{r}) - {\bar{\cal B}} \,R^{(0)}_{\ell m\omega} (\tilde{r}) \Big\}, \label{EqTS11}
\end{eqnarray}
where $A^{(1)}_0 =A^{(1)} _{nn\,0}+A ^{(1)}_{{\overline m}n\,0}+
A^{(1)} _{{\overline m}{\overline m}\,0},$ $
 A^{(1)}_1 =A^{(1)} _{{\overline m}n\,1}+A ^{(1)}_{{\overline m}{\overline m}\,1},$  and $ A^{(1)}_2 =A^{(1)} _{{\overline m}{\overline m}\,2}$.
For the circular orbits, we can take $r(t)=r_0$, $\theta(t)=\theta_0$, and $\varphi(t)=\Omega\, t$, where $\Omega$ is the angular velocity. Noting that $ \tilde Z^{(1)}_{\ell m\omega}= Z^{(1)}_{\ell m\omega}\delta(\omega-\omega_n)$ \cite{Jing}, we then find that Eq. (\ref{EqTS11}) can be expressed as
\begin{eqnarray}
\tilde Z ^{(1)}_{\ell m\omega}&=&
\frac{\pi m_0}{i\omega B^{\rm inc}_{\ell m\omega}}\Bigg\{ G \Biggl[ A ^{(1)}_0
\frac{R^{\rm in(0)}_{\ell m\omega}(r)}{r^3 F_{40}}- A ^{(1)}_1
\Big(\frac{R^{\rm in(0)}_{\ell m\omega}(r)}{r^3 F_{40}}\Big)' + A ^{(1)}_2 \Big(\frac{R^{\rm in(0)}_{\ell m\omega}(r)}{r^3 F_{40}}\Big)'' \Biggr]_{r_0,\theta_0}\nonumber \\ &+& B^{(1)}_{eff}\Bigg\}\delta(\omega-\omega_n),
 \label{ZZZSchS}
\end{eqnarray}
where
\begin{align}
 B^{(1)}_{eff}&\delta(\omega-\omega_n)= \nonumber \\
 &\frac{1}{ 2\pi m_0}
\int^{\infty}_{r_+}d\tilde{r} \frac{
R^{\rm in(0)}_{\ell m\omega}(\tilde{r} ) }{ \tilde{r} ^3 F_{40} (\Delta^0_{r})^2}\Bigg\{ {\cal Q}_{\ell} \Big[\frac{i \Delta^0_{\tilde{r}}}{\tilde{r}^2}\big(3+\tilde{r}^2 F_{11}(\tilde{r})\big)\frac{d}{d \tilde{r}} +U(\ell-1)\Big] R^{(0)}_{\ell-1 m\omega}(\tilde{r} ) \, \nonumber \\
&+ {\cal Q}_{\ell+1} \Big[\frac{i \Delta^0_{\tilde{r}}}{\tilde{r}^2}\big(3+r^2 F_{11}(\tilde{r})\big)\frac{d}{d \tilde{r}} +U(\ell+1)\Big] R^{(0)}_{\ell+1 m\omega}(\tilde{r} ) \, - \, {\bar{\cal B}} R^{(0)}_{\ell m\omega} (\tilde{r} ) \Bigg\}.
 \label{ZZZSchSS}
\end{align}
Therefore, the first-order formal solution of $\psi_4^B$ is described by
\begin{eqnarray}
\psi^{B(1)}_4&=&\frac{1}{ r}\sum_{\ell m n}
\frac{\pi m_0}{i\omega_n B^{\rm inc}_{\ell m\omega_n}}
\Bigg\{ G \Biggl[ A ^{(1)}_0
\frac{R^{\rm in(0)}_{\ell m\omega}(r)}{r^3 F_{40}}- A ^{(1)}_1
\Big(\frac{R^{\rm in(0)}_{\ell m\omega}(r)}{r^3 F_{40}}\Big)' + A ^{(1)}_2 \Big(\frac{R^{\rm in(0)}_{\ell m\omega}(r)}{r^3 F_{40}}\Big)'' \Biggr]_{r_0,\theta_0}\nonumber \\ &+& B^{(1)}_{eff}\Bigg\}\frac{{}_{-2}Y_{\ell m} }{ \sqrt{2\pi}}
e^{i\omega_n(r^*-t)+im\varphi}.
\label{psi412}
\end{eqnarray}

\vspace{0.5cm}
 {\bf 3-3. Solution of $\psi^B_{4}$ in the effective spacetime}

Equations (\ref{psi41}) and (\ref{psi412}) show that the solution of $\psi^B_{4}$ up to the first order of the rotational parameter $a$ in the effective spacetime can be expressed as
\begin{eqnarray}
\psi^{B}_4&=&\frac{1}{ r}\sum_{\ell m n}
\frac{\pi G m_0}{i\omega_n B^{\rm inc}_{\ell m\omega_n}}
\Bigg\{ \Biggl[ (A^{(0)}_0+a A^{(1)}_0 )
\frac{R^{\rm in(0)}_{\ell m\omega_n}(r)}{r^3 F_{40}}- (A^{(0)}_1 +a A^{(1)}_1)\Big(\frac{R^{\rm in(0)}_{\ell m\omega_n}(r)}{r^3 F_{40}}\Big)' \nonumber \\ &+& (A^{(0)}_2+aA^{(1)}_2)\Big(\frac{R^{\rm in(0)}_{\ell m\omega_n}(r)}{r^3 F_{40}}\Big)'' \Biggr]_{r_0,\theta_0}
 +\frac{a}{G} B^{(1)}_{eff}\Bigg\}\frac{{}_{-2}Y_{\ell m} }{ \sqrt{2\pi}}
e^{i\omega_n(r^*-t)+im\varphi}.
\label{psi4T}
\end{eqnarray}

\subsection{Radiation reaction force and waveform for ``plus" and ``cross" modes of the gravitational wave }

The reduced RRF is described by \cite{Buonanno2006}
\begin{eqnarray} \hat{\bm{\mathcal{F}}}=\frac{1}{\nu M_0 \Omega |\vvr\times
   \vp|}\frac{dE}{dt}\vp,
\end{eqnarray}
where $\Omega = |\vvr\times\dot{\vvr}|/r^2$ is the dimensionless orbital frequency and $dE/dt$ is the energy flux of the gravitational radiation.
For the quasicircular cases without precession, noting that $|\vvr\times \vp|\approx p_\varphi $, we find that the reduced RRF that appeared in the Hamiltonian equation (\ref{HEq}) can be expressed explicitly as
\begin{eqnarray}
\hat{\bm{\mathcal{F}}}&=&\frac{1}{\nu M_0 \Omega}\sum_{\scriptstyle \ell=2}^{\infty}\sum_{m=1}^\ell
\dfrac{2 \pi m_0^2 }{ G\omega_n^4}\Bigg| \frac{G}{B^{\rm inc}_{\ell m\omega_n}}
\Bigg\{ \Biggl[ (A^{(0)}_0+a A^{(1)}_0 )
\frac{R^{\rm in(0)}_{\ell m\omega_n}(r)}{r^3 F_{40}}- (A^{(0)}_1 +a A^{(1)}_1)\Big(\frac{R^{\rm in(0)}_{\ell m\omega_n}(r)}{r^3 F_{40}}\Big)' \nonumber \\ &+& (A^{(0)}_2+aA^{(1)}_2)\Big(\frac{R^{\rm in(0)}_{\ell m\omega_n}(r)}{r^3 F_{40}}\Big)'' \Biggr]_{r_0,\theta_0}
 +\frac{a}{G} B^{(1)}_{eff}\Bigg\}\Bigg|^2\frac{\vp}{p_{\varphi}},
 \label{FFdE1}
\end{eqnarray}
indicating that the reduced RRF is constructed in terms of the effective spacetime.

On the other hand, it is well known that the ``plus" and ``cross" modes of the gravitational wave can be expressed in terms of spin-weighted $s=-2$ spherical harmonics~\cite{Kidder}:
\begin{align}
h_{+}-i h_{\times}=\sum_{l=2}^{\infty} \sum_{m=-l}^{l} h^{lm}\frac{ \ _{-2}Y^{lm}(\theta, \varphi)}{\sqrt{2\pi}} .\label{hh}
\end{align}
Thus, by comparing Eq.~(\ref{hh}) with the solution of $\psi_4^B$, we can read off the waveform easily, which is given by
\begin{eqnarray}
h^{l m}&=&\frac{1}{ r }
\frac{2 \pi G m_0 }{i\omega_n^3 B^{\rm inc}_{\ell m\omega_n}}
\Bigg\{ \Biggl[ (A^{(0)}_0+a A^{(1)}_0 )
\frac{R^{\rm in(0)}_{\ell m\omega_n}(r)}{r^3 F_{40}}- (A^{(0)}_1 +a A^{(1)}_1)\Big(\frac{R^{\rm in(0)}_{\ell m\omega_n}(r)}{r^3 F_{40}}\Big)' \nonumber \\ &+& (A^{(0)}_2+aA^{(1)}_2)\Big(\frac{R^{\rm in(0)}_{\ell m\omega_n}(r)}{r^3 F_{40}}\Big)'' \Biggr]_{r_0,\theta_0}
 +\frac{a}{G} B^{(1)}_{eff}\Bigg\}e^{i\omega_n(r^*-t)}.
\label{hform}
\end{eqnarray}
 It is clear that the waveform is also constructed in terms of the effective rotating spacetime. Based on $h^{lm}$ described by Eq.~(\ref{hform}) and using Damour--Nagar's proposals \cite{Damour2007,Damour2009,Pan2,Pan3}, we can then construct an improved inspiral--plunge mode, i.e., an improved multipolar waveform can be built as a product of
 the leading Minkowskian order term $h_{\ell m}^{(M,\epsilon_p)}$, the relativistic conserved energy or angular momentum of the effective moving source $\hat{S}_{\rm eff}^{(\epsilon_p)}$, an infinite number of leading logarithms that enter the tail effect $T_{\ell m}$, an additional phase correction $e^{i\delta_{\ell m}}$, the remaining PM effects $f_{\ell m}$, and a non-quasicircular effect $N_{lm}$.

The above discussions show that the improved reduced Hamiltonian (\ref{Heob}), reduced RRF (\ref{FFdE1}), and waveform (\ref{hform}) for the ``plus" and ``cross" modes of the gravitational wave, which are based on the same physical model, constitute a self-consistent EOB theory for the spin binaries based on the PM approximation.

\section{ Conclusions and discussions}

From a practical point of view, it is well known that the general gravitational waveform template should be constructed for the spin binaries. Therefore, as a theory to construct the gravitational waveform template, the self-consistent EOB theory should be set up for the spinning black hole binaries. In this paper, we extend the study on the SCEOB theory for the real spinless two-body system \cite{Jing} to the case of the real spin two-body system.

Based on the effective metric (\ref{Mmetric}) for a real spinless two-body system in the post-Minkowskian approximation obtained in Ref. \cite{Jing}, we used Damour's or Barausse's approach \cite{Barausse,Damour} to find the effective rotating metric (\ref{effmetric}) for a real spin two-body system. The effective rotating metric is still type $D$. In principle, it can be applicable to any post-Minkowskian order. By means of the effective metric (\ref{effmetric}), the energy map (\ref{EMap}), and the approach of constructing the EOB Hamiltonian for the spinning black hole binaries presented by Damour and Barausse et al.,~\cite{Damour,DamourH,BarausseH,BarausseH1}, we then obtain an improved reduced EOB Hamiltonian (\ref{Heob}) of the EOB theory for the real spin two-body system at a linear order of the particle's spin which appears in the Hamilton equations (\ref{HEq}).

To work out the reduced RRF and the waveform for the ``plus" and ``cross" modes of the gravitational wave generated by a coalescing compact object binary system,
we first used the gauge transform property of tetrad components of the perturbed Weyl tensors to find the decoupled equation (\ref{EqT}) of the null tetrad component of the gravitational perturbed Weyl tensor $\psi^B_{4}$ in the effective rotating spacetime.
Next, noting that the decoupled equation (\ref{EqT}) does not appear to be directly separable, we find that we can separate the equation for the radial and angular parts in the slowly rotating background, which is close enough to the spherical symmetry, and derive the explicitly radial equation (\ref{EqaT}) up to the first order of the rotational parameter $a$. However, it should be noted that, in principle, the method can be extended to any order of the rotational parameter $a$.
Using the energy-momentum tensor (\ref{Tgen}) for a particle that orbits around a massive black hole described by the effective rotating metric,
we then note that the radial equation can be solved order by order in the rotational parameter $a$ and present a formal solution (\ref{psi4T}) of $\psi^B_{4}$ up to the first order of the rotational parameter $a$ by means of the Green function.
Finally, with the help of the null tetrad component of the perturbed Weyl tensor $\psi^B_{4}$, we present the formal expressions of the reduced RRF (\ref{FFdE1}) and the waveform (\ref{hform}) for ``plus" and ``cross" modes of the gravitational wave in the effective rotating spacetime.

It is interesting to note that the improved reduced EOB Hamiltonian (\ref{Heob}), the reduced RRF (\ref{FFdE1}), and the waveform (\ref{hform}) for the ``plus" and ``cross" modes of the gravitational wave, which are obtained in the same physical model, constitute a self-consistent EOB theory for the spin black hole binaries based on the PM approximation.

\vspace{0.3cm}
\acknowledgments
{ We would like to thank professors S. Chen and Q. Pan  for useful discussions on the manuscript. This work was supported by the Grant of NSFC Nos. 12035005 and 12122504, and National Key Research and Development  Program of China No. 2020YFC2201400.}  

\end{document}